\documentclass[fleqn,usenatbib]{mnras}
\usepackage{url,times,graphicx,amsmath,amsfonts,amssymb,color,epsfig,epstopdf}
\usepackage[T1]{fontenc}
\DeclareRobustCommand{\VAN}[3]{#2}
\let\VANthebibliography\thebibliography
\def\thebibliography{\DeclareRobustCommand{\VAN}[3]{##3}\VANthebibliography}

\usepackage[british]{babel}
\usepackage[inline]{enumitem}
\usepackage{multirow}
\usepackage[slantedGreek]{newtxmath}
\usepackage{newtxtext}
\usepackage{tabularx}
\usepackage{catchfile}
\usepackage{soul}

\usepackage{soul} 
\usepackage{amsmath}
\usepackage{newtxtext}
\usepackage[slantedGreek]{newtxmath}
\usepackage{xifthen}
\usepackage{xspace}

\usepackage{ulem}
\usepackage[dvipsnames]{xcolor}
\definecolor{harvestgold}{rgb}{0.85, 0.57, 0.0}
\definecolor{palatinate}{RGB}{174, 0, 201}
\definecolor{darkblue}{RGB}{0, 0, 139}
\definecolor{burntpink}{RGB}{220, 0, 190}
\definecolor{burntred}{RGB}{186, 41, 0}
\definecolor{lightcyan}{RGB}{22, 214, 240}
\definecolor{limegreen}{RGB}{191, 204, 43}

\usepackage[hyperref]{ntheorem}
\newtheorem{quest}{Question}[section]

\newtheorem{disc}{Discussion}[section]

\newcommand{\orcid}[1]{\href{https://orcid.org/#1}{\,\includegraphics[height=\fontcharht\font`\B]{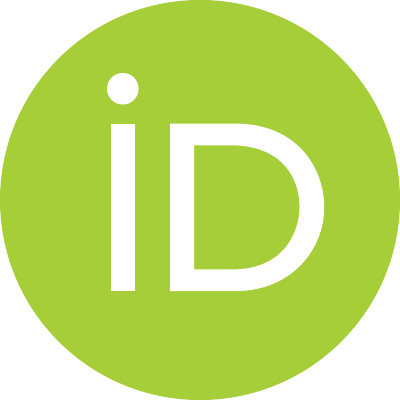}}}
\newcommand{\github}[1]{\href{https://github.com/#1}{\includegraphics[height=\fontcharht\font`\B]{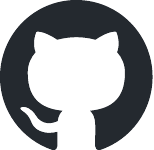} \nolinkurl{#1}}}

\newcommand{\percent}[1]{\ensuremath{#1}~per cent}

\newcolumntype{L}{>{\raggedright\arraybackslash}X}
\newcolumntype{Y}{>{\centering\arraybackslash}X}
\newcolumntype{R}{>{\raggedleft\arraybackslash}X}

\newcommand{\WMAPsevenp}{\ensuremath{{\Ho[{\kms[70.4]\pMpc{}}]},\,\allowbreak {\Omegaup_{\rm M}=0.272},\,\allowbreak {\Omegaup_\Lambdaup=0.728},\,\allowbreak {n_s=0.967},\, \sigma_8=0.81}}
\newcommand{\PLANCKeighteenp}{\ensuremath{{\Ho[{\kms[67.4\pm0.5]\pMpc{}}]},\,\allowbreak {\Omegaup_{\rm M}=0.315\pm0.007},\,\allowbreak {\Omegaup_\Lambdaup=0.685\pm0.007},\,\allowbreak {n_s=0.965\pm0.004},\, \sigma_8=0.811\pm0.006}}

\makeatletter
\newcommand{\phantomlabel}[2]{
    \protected@write\@auxout{}{
        \string\newlabel{#2}{
            {\@currentlabel, #1}{\thepage}
            {\@currentlabel, #1}{#2}{}
        }
    }%
    \hypertarget{#2}{}%
}
\makeatother

\newcommand{\appref}[1]{Appendix~\ref{#1}}

\newcommand{\secref}[1]{Section~\ref{#1}}
\newcommand{\secrefs}[2]{Sections~\ref{#1} and \ref{#2}}
\newcommand{\figref}[1]{Fig.~\ref{#1}}
\newcommand{\figrefs}[2]{Figs~\ref{#1}~and~\ref{#2}}
\newcommand{\eqnref}[1]{equation~(\ref{#1})}
\newcommand{\eqnrefp}[1]{equation~\ref{#1}}

\newcommand{\extapp}[1]{appendix~#1}

\newcommand{\extfig}[1]{fig.~\ensuremath{#1}}
\newcommand{\extsec}[1]{section~\ensuremath{#1}}
\newcommand{\exttab}[1]{table~\ensuremath{#1}}

\defcitealias{tollerud_hundreds_2008}{T08}
\defcitealias{koposov_luminosity_2008}{K08}
\defcitealias{walsh_invisibles_2009}{W09}
\defcitealias{jethwa_magellanic_2016}{J16}
\defcitealias{newton_total_2018}{N18}
\defcitealias{lacey_unified_2016}{L16}

\newcommand{\SigmaFunc}[1]{\ensuremath{\sigma\!\left(#1\right)}}

\newcommand{\bigO}[1]{\ensuremath{\mathop{}\mathopen{}\mathcal{O}\mathopen{}\left(#1\right)}}
\newcommand{\dv}[1]{\ensuremath{\mathrm{d}#1}} 
\newcommand{\fdv}[2]{\ensuremath{\frac{\dv{#1}}{\dv{#2}}}} 

\newcommand{\MsConstraint}[1]{\ensuremath{\Ms{}\leq\keV[#1]}}

\newcommand{\Fermi}{{\textit{Fermi}}}
\newcommand{\FermiLAT}{{\Fermi{}--LAT}}
\newcommand{\Gaia}{{\textit{Gaia}}}
\newcommand{\Hitomi}{{\textit{Hitomi}}}

\newcommand{\XRISM}{{\textit{XRISM}}}

\newcommand{\Apostle}{{\sc APOSTLE}}

\newcommand{\COCO}{\mbox{{\sc COCO}}}

\newcommand{\Millennium}{{Millennium}}

\newcommand{\Galform}{{\sc galform}}

\newcommand{\astropy}{{\sc Astropy}}

\newcommand{\numpy}{{\sc numpy}}
\newcommand{\python}{{\sc python}}

\newcommand{\scipy}{{\sc scipy}}
\newcommand{\matplotlib}{{\sc matplotlib}}

\newcommand{\CDM}{{cold dark matter}}

\newcommand{\DM}{{dark matter}}
\newcommand{\EPS}{{Extended Press--Schechter}}

\newcommand{\hSpher}{\ensuremath{s_{\rm half-max}}}

\newcommand{\LCDM}{{$\Lambdaup$CDM}}
\newcommand{\LMC}{{LMC}}
\newcommand{\Lyman}[1]{Lyman~\ensuremath{#1}}
\newcommand{\M}[1]{\ensuremath{M_{#1}}}

\newcommand{\MW}{Milky Way}
\newcommand{\MDeltaC}[2][200]{%
  \ifthenelse{\isempty{#2}}%
    {\ensuremath{\M{#1}}}
    {\ensuremath{\M{#1}^{{\rm #2}}}}
  }

\newcommand{\Nbody}{\textit{N}--body}
\newcommand{\Nsat}[1][]{%
  \ifthenelse{\isempty{#1}}%
    {\ensuremath{N_{\rm sat}}}
    {\ensuremath{N^{{\rm #1}}_{\rm sat}}}
  }
\newcommand{\Nsub}[1][]{%
  \ifthenelse{\isempty{#1}}%
    {\ensuremath{N_{\rm sub}}}
    {\ensuremath{N^{#1}_{\rm sub}}}
  }
\newcommand{\R}[1]{\ensuremath{R_{#1}}}
\newcommand{\RNFW}{\R{200}}
\newcommand{\RDeltaC}[1][]{%
  \ifthenelse{\isempty{#1}}%
    {\ensuremath{\R{200}}}
    {\ensuremath{\R{200}^{{\rm #1}}}}
  }

\newcommand{\WDM}{{WDM}}

\newcommand{\WMAP}{{WMAP}}

\newcommand{\unit}[1]{\ensuremath{\mathrm{\,#1}}\xspace}
\newcommand{\unitlogicnospace}[2]{%
  \ifthenelse{\isempty{#1}}%
    {\unit{#2}}
    {\ensuremath{{{#1}\unit{#2}}}}
  }
\newcommand{\unitlogicspace}[2]{%
  \ifthenelse{\isempty{#1}}%
    {\unit{#2}}
    {\ensuremath{{{#1}\, \unit{#2}}}}
  }

\newcommand{\Msun}[1][]{\unitlogicspace{#1}{M_\odot}}
\newcommand{\Msunh}[1][]{\unitlogicspace{#1}{\mathit{h}^{-1}\Msun{}}}

\newcommand{\pMpc}[1][]{\unitlogicspace{#1}{Mpc^{-1}}}
\newcommand{\keV}[1][]{\unitlogicspace{#1}{keV}}
\newcommand{\MeV}[1][]{\unitlogicspace{#1}{MeV}}
\newcommand{\GeV}[1][]{\unitlogicspace{#1}{GeV}}

\newcommand{\kms}[1][]{\unitlogicspace{#1}{km\, s^{-1}}}

\newcommand{\variablelogicspace}[2]{%
  \ifthenelse{\isempty{#2}}%
    {\ensuremath{#1}}
    {\ensuremath{{{#1}={#2}}}}
  }
\renewcommand{\d}[1]{\ensuremath{\operatorname{d}\!{#1}}}

\newcommand{\fviable}[1][]{\variablelogicspace{f_{\rm v}}{#1}}

\newcommand{\sigmaI}[1][]{\variablelogicspace{\upsigma_{\rm I}}{#1}}
\newcommand{\sigmaPoisson}[1][]{\variablelogicspace{\upsigma_{\rm Poisson}}{#1}}

\newcommand{\z}[1][]{\variablelogicspace{z}{#1}}
\newcommand{\zreion}[1][]{\variablelogicspace{z_\mathrm{reion}}{#1}}

\newcommand{\Ho}[1][]{\variablelogicspace{H_0}{#1}}

\newcommand{\Mhm}[1][]{\variablelogicspace{\M{\rm hm}}{#1}}

\newcommand{\MNFW}[1][]{\variablelogicspace{M_{200}}{#1}}
\newcommand{\MMW}[2][200]{\variablelogicspace{\MDeltaC[#1]{MW}}{#2}}

\newcommand{\MV}[1][]{\variablelogicspace{M_\mathrm{V}}{#1}}

\newcommand{\NsatNum}[2][]{\variablelogicspace{\Nsat[#1]}{#2}}

\newcommand{\Vcirc}[1][]{\variablelogicspace{V_{\rm circ}}{#1}}
\newcommand{\Vcut}[1][]{\variablelogicspace{V_{\rm cut}}{#1}}

\newcommand{\MSM}{{\ensuremath{\nu}MSM}}
\newcommand{\Ms}[1][]{\variablelogicspace{M_\mathrm{s}}{#1}}
\newcommand{\mixAng}{\ensuremath{\theta_{\rm e}}}
\graphicspath{{Figures/}}

\title[Constraining \MSM{} using Milky Way satellites]{Constraints on the properties of \MSM{} dark matter using the satellite galaxies of the Milky Way}

\author[O. Newton et al.]{Oliver Newton\orcid{0000-0002-2769-9507},$^{1}$\thanks{E-mail: onewton@cft.edu.pl}
Mark R.~Lovell\orcid{0000-0001-5609-514X},$^{2}$
Carlos S.~Frenk\orcid{0000-0002-2338-716X},$^2$
Adrian Jenkins\orcid{0000-0003-4389-2232},$^2$
John C.~Helly\orcid{0000-0002-0647-4755},$^2$
\newauthor
Shaun Cole\orcid{0000-0002-5954-7903}$^2$
and Andrew J.~Benson\orcid{0000-0001-5501-6008}$^3$
\\
$^1$Center for Theoretical Physics, Polish Academy of Sciences, al. Lotnik\'{o}w 32/46 Warsaw, Poland\\
$^2$Institute for Computational Cosmology, Durham University, South Road, Durham, DH1 3LE, UK\\
$^3$Carnegie Observatories, 813 Santa Barbara Street, Pasadena, CA 91101, U.S.A
\vspace{-10pt}}

\date{Accepted XXX. Received YYY; in original form ZZZ
\vspace{-10pt}}

\pubyear{2024}

\begin{document}
\label{firstpage}
\pagerange{\pageref{firstpage}--\pageref{lastpage}}
\maketitle

\begin{abstract}
Low-mass galaxies provide a powerful tool with which to investigate departures from the standard cosmological paradigm in models that suppress the abundance of small dark matter structures.
One of the simplest metrics that can be used to compare different models is the abundance of satellite galaxies in the \MW{}.
Viable \DM{} models must produce enough substructure to host the observed number of Galactic satellites.
Here, we scrutinize the predictions of the neutrino Minimal Standard Model~(\MSM{}), a well-motivated extension of the Standard Model of particle physics in which the production of sterile neutrino \DM{} is resonantly enhanced by a lepton asymmetry in the primordial plasma.
This process enables the model to evade current constraints associated with non-resonantly produced \DM{}.
Independently of assumptions about galaxy formation physics we rule out, with at least \percent{95} confidence, all parameterizations of the \MSM{} with sterile neutrino rest mass, \MsConstraint{1.4}.
Incorporating physically motivated prescriptions of baryonic processes and modelling the effects of reionization strengthen our constraints, and we exclude all \MSM{} parameterizations with \MsConstraint{4}.
Unlike other literature, our fiducial constraints do not rule out the putative \keV[3.55] X-ray line, if it is indeed produced by the decay of a sterile neutrino; however, some of the most favoured parameter space is excluded.
If the \MW{} satellite count is higher than we assume, or if the \MW{} halo is less massive than \MMW{\Msun[{8\times10^{11}}]}, we rule out the \MSM{} as the origin of the \keV[3.55] excess.
In contrast with other work, we find that the constraints from satellite counts are substantially weaker than those reported from X-ray non-detections.
\end{abstract}

\begin{keywords}
dark matter -- elementary particles -- Galaxy: halo -- galaxies: dwarf -- galaxies:formation
\vspace{-10pt}
\end{keywords}


\section{Introduction}
\label{sec:Introduction}
Understanding the nature of dark matter is one of the greatest outstanding challenges in contemporary physics.
In the prevailing cosmological framework known as $\upLambda+$\CDM{}~(\LCDM{}) most of the \DM{} is typically composed of weakly interacting, massive fundamental particles (WIMPs)
exhibiting negligible thermal velocities at early times \citep{peebles_large-scale_1982,davis_evolution_1985,bardeen_statistics_1986}.
Although \LCDM{} successfully predicts and replicates various observable features of the Universe, terrestrial direct detection \citep[e.g.][]{supercdms_collaboration_improved_2015,cresst_collaboration_first_2019,deap_collaboration_search_2019,pico_collaboration_dark_2019,pandax-4t_collaboration_dark_2021,aalbers_first_2023,abe_direct_2023,agnes_search_2023,aprile_first_2023},
indirect detection \citep[][also see \citealp{murgia_fermilat_2020} for a comprehensive review of \Fermi{} Large Area Telescope~(\FermiLAT{}) searches for a $\gamma$-ray excess]{albert_results_2017,hess_collaboration_search_2018,di_mauro_search_2019,the_super-kamiokande_collaboration_indirect_2020,regis_emu_2021,icecube_collaboration_search_2022,di_mauro_constraining_2023,mcdaniel_legacy_2024,yin_first_2024},
and particle collider searches \citep{atlas_collaboration_search_2021,atlas_collaboration_search_2021-1,cms_collaboration_search_2021,tumasyan_search_2022,tumasyan_search_2022-1,aad_combination_2023,tumasyan_search_2023-1}
have yet to establish
that weakly interacting massive \DM{} particles exist.
This has encouraged closer examination of alternative \DM{} models that are well motivated in the particle physics context and which may be compatible with the successes achieved on
astrophysical
scales by \LCDM{}.

Astrophysical observations serve as critical benchmarks against which to evaluate various aspects of these alternative models.
In particular, low-mass galaxies provide valuable
insights into the physics governing the onset of structure formation and its subsequent evolution. It is within this regime that departures from the prevailing paradigm are likely to manifest.
Models in which the \DM{} is relativistic in the early Universe and undergoes free-streaming on scales that suppress the formation of low-mass haloes may struggle to account for the observed abundance of Milky Way satellite galaxies. This has proven to be an effective means of imposing stringent constraints on several `warm' \DM{} models \citep[\WDM{}, e.g.][]{kennedy_constraining_2014,lovell_satellite_2016,enzi_joint_2021,nadler_constraints_2021,newton_constraints_2021,dekker_warm_2022}.
In this study we adopt this approach, described in detail in \citet{newton_constraints_2021}, to impose constraints on the parameter space of one well-motivated model: the neutrino Minimal Standard Model~(\MSM{}).

The \MSM{}, a minimal extension of the Standard Model of particle physics,
introduces
three
`sterile' neutrinos alongside the family of three standard model `active' neutrinos \citep{shi_new_1999,asaka_msm_2005-1,laine_sterile_2008}.
The sterile neutrinos are produced by the oscillation of the left-chiral active neutrinos into right-chiral states, characterized by the mixing angle, $\theta$. The mixing of the active and sterile neutrino states produces two distinct sets of mass eigenstates wherein the masses in one set increase as the neutrinos in the other set become less massive \citep{minkowski_e_1977,gell-mann_complex_1979,barbieri_neutrino_1980,mohapatra_neutrino_1980,yanagida_horizontal_1980}.
This elegant mechanism offers a plausible explanation for the extremely small masses of the active neutrinos of the Standard Model.
The high mass eigenstates are ascribed to the sterile neutrinos: one with a mass \bigO{\keV{}} that is long-lived and exhibits relativistic behaviour up to nine years after the Big Bang \citep{lovell_anticipating_2023}, and two others with masses in the \GeV{} scale that are unstable and extremely short-lived (\citealp{adhikari_white_2017}; for a detailed review see \citealp{abazajian_light_2012}).
Their decay into anti-leptons introduces an asymmetry in the lepton abundance \citep{fukugita_barygenesis_1986} that is partially reprocessed into a baryon asymmetry during baryogenesis. This provides a natural explanation for the observed asymmetry of matter to anti-matter in the Universe \citep{shaposhnikov_$upnu$msm_2008,canetti_matter_2012,canetti_sterile_2013}.
A recent comprehensive description of the production mechanisms underpinning the \MSM{} in an astrophysical context is available in \citet{lovell_anticipating_2023}.

As active neutrinos propagate through a medium their oscillation parameters may diverge from those \textit{in vacuo} because of interactions with electrons in the medium. This phenomenon is known as the {Mikheyev--Smirnov--Wolfenstein}~(MSW) effect \citep{wolfenstein_neutrino_1978,mikheyev_resonance_1985}.
An analogous mechanism within the primordial plasma affects the mixing of the active and sterile neutrino states in the \MSM{}. This is controlled by the size of the lepton asymmetries in different particle species, which can resonantly enhance the production of sterile neutrinos by increasing the effective mixing angle, \mixAng{} \citep{laine_sterile_2008}.
Through this process, the \MSM{} may generate sufficient numbers of long-lived \keV{} scale sterile neutrinos to be consistent with the measured abundance of \DM{}.
Since the sterile neutrinos decouple from the primordial plasma before the onset of Big Bang nucleosynthesis they do not affect the primordial abundance of chemical elements. Consequently, astrophysical measurements of the elemental abundances, which have been used to constrain the properties of thermally produced \WDM{} candidates, cannot be applied to constrain the \MSM{} parameter space \citep{sabti_refined_2020,an_what_2022}.

Efforts to ascertain the nature of the \DM{} intensified in response to astrophysical spectroscopic observations revealing a possible spectral line at \keV[3.55] in systems containing a significant \DM{} component \citep{bulbul_detection_2014,boyarsky_unidentified_2014,cappelluti_searching_2018,hofmann_71_2019}.
This has elicited considerable interest, given the absence of a known astrophysical origin for a spectral line at this energy.
While charge exchange between sulphur ions and a neutral medium could provide a plausible baryonic explanation \citep{gu_novel_2015,shah_laboratory_2016}, it cannot be confirmed with the resolution available in current instruments.\footnote{The instruments onboard \Hitomi{} filled this gap in capability prior to its rapid unscheduled disassembly. Unfortunately, the data collected were insufficient to draw definitive conclusions regarding charge exchange as the origin of the X-ray excess. The shortfall in spectral resolution will be rectified when the newly launched \XRISM{} satellite starts taking performance verification data.} Moreover, it may be unable to account for all of the observed flux in any case \citep{cappelluti_searching_2018}.
The analysis of the excess is complicated further by the choice of assumptions embedded in the modelling of the astrophysical background \citep[][also see \citealp{boyarsky_sterile_2019} for a recent review]{dessert_was_2024}.

A non-baryonic explanation for the excess could be provided by the dark sector. Interpreting the line as the decay product of a \DM{} particle would place the rest mass of the parent particle within the range expected of a sterile neutrino candidate responsible for the entire \DM{} component of the cosmic matter density.
Follow-up observations of the spectral feature in the Galactic centre \citep{boyarsky_checking_2015,perez_almost_2017,hofmann_71_2019}, the Perseus cluster \citep{urban_suzaku_2015,franse_radial_2016}, and in other galaxy clusters \citep[][but see also \citealp{hofmann_7.1_2016}]{bulbul_searching_2016} have been carried out in attempts to improve the significance of the detection and elucidate the origin of the excess, with limited success. These plausible detections are challenging to reconcile with the non-detections reported by other observational studies of stacked galaxy spectra \citep{anderson_non-detection_2015}, the Galactic centre \citep{riemer-sorensen_constraints_2016}, the Draco dwarf galaxy \citep{ruchayskiy_searching_2016,sonbas_x-ray_2016}, the X-ray background \citep{sekiya_search_2016}, and Perseus \citep{aharonian_hitomi_2017,tamura_x-ray_2019}.
The origin of the \keV[3.55] excess thus remains unclear, and this emphasizes the need for higher resolution measurements of the soft X-ray spectrum in \DM{}-dominated systems. This will be achievable with the JAXA \XRISM{} mission that launched in 2023 September that is now in the final stages of commissioning \citep{tashiro_concept_2018,lovell_signal_2019,lovell_simulating_2019,terada_detailed_2021,dessert_resurrecting_2024}.

In this work, we impose constraints on the parameter space of resonantly produced \MSM{} models by comparing the predicted abundance of low-mass galaxies in \MW{} mass \DM{} haloes with the inferred total population of \MW{} satellite galaxies.
In \secref{sec:Methods}, we summarize the method as applied to the \MSM{} and calibrate it using \DM{}-only \Nbody{} simulations of \MSM{} models. We present our constraints on the \MSM{} parameter space in \secref{sec:Results} and compare them with related studies. We discuss our results in \secref{sec:Discussion} and present concluding remarks in \secref{sec:Conclusions}.

\section{Methods}
\label{sec:Methods}
\citet{newton_constraints_2021} described an approach to constrain the parameter space of thermal relic \WDM{} models by comparing
predictions of the abundance of substructure in \MW{} mass \DM{} haloes with the
total number of satellite galaxies estimated from observations of the \MW{} halo.
Here, we apply the same technique to constrain the viable combinations of the sterile neutrino rest mass, \Ms{}, and the resonantly enhanced effective mixing angle, \mixAng{}, in the \MSM{}.
We make predictions of the abundance of substructure in different parameterizations of the \MSM{} using the \EPS{} formalism, first computing the \MSM{} momentum distributions and matter power spectra in \secref{sec:Methods:ComputePk}, then using \EPS{} theory to predict the subhalo counts;
we describe and calibrate the \EPS{} formalism using gravity only cosmological simulations of \MW{} mass haloes in \secref{sec:Methods:Calibrate_EPS}.
In \secref{sec:Methods:Calculate_acceptance_probability}, we summarize the approach we take to evaluate the viability of each parameterization of the model.

\subsection{Computing the $\nu$MSM sterile neutrino power spectra}
\label{sec:Methods:ComputePk}

The primary goal of this study is to predict the abundance of satellites of galaxies like the \MW{} as a function of the \DM{} model parameters. The first step is to compute the momentum distribution of the \DM{} at an energy of \MeV[1], or approximately one second after the  Big Bang, which is then input to a Boltzmann solver code that calculates the linear matter power spectrum, $P\!\left(k\right)$. We compute momentum distributions for an array of 36 values of \Ms{} and 31 values of the lepton asymmetry parameter, $L_6,$ using the code of \citet{laine_sterile_2008}. The $P\!\left(k\right)$ are then calculated with a modified version of the {\sc camb} Boltzmann solver \citep{lewis_efficient_2000}.

One of the key uncertainties in our results will be the accuracy of the momentum distribution calculation. The \citet{laine_sterile_2008} algorithm has been superseded by that of \citet{ghiglieri_improved_2015}, with minor modifications to the free-streaming length. A further study by \citet{venumadhav_sterile_2016} produced results that differ significantly from those of \citet{ghiglieri_improved_2015}: the former predicts a cut-off scale some three times larger than the latter at fixed sterile neutrino mass and effective mixing angle \citep{lovell_anticipating_2023}. We will discuss this discrepancy further in \secref{sec:Results:nuMSM_constraints:Num_MW_sats}.

\subsection{Calibrating the \EPS{} formalism with numerical simulations}
\label{sec:Methods:Calibrate_EPS}

In the \EPS{} formalism the primordial matter density field is filtered using a window function to identify regions that are dense enough to collapse into virialized haloes by the present day \citep{press_formation_1974,bond_excursion_1991,bower_evolution_1991,lacey_merger_1993,parkinson_generating_2008}.
The abundance of structure at different mass scales can be calculated from this using the conditional halo mass function, $\d{N_{\rm cond}}\, /\, \d{\, \ln M}$.
This describes the fraction of the mass in some halo at time, $t_1$, that was in collapsed progenitors of a given mass, \textit{M}, at some earlier time, $t_2$.
The conditional halo mass function depends on the primordial linear matter power spectrum, and its functional form and the details of its derivation are provided in \citet{schneider_structure_2015}.

The subhalo mass function is computed by integrating the conditional halo mass function over the redshift-dependent overdensity collapse threshold of a given progenitor, $\delta_c\!\left(z\right)$. It is given by
\begin{equation}
    \fdv{\Nsub{}}{\ln M} = \frac{1}{N_{\rm norm}} \int_{\delta_c\!\left(0\right)}^\infty
    \fdv{N_{\rm cond}}{\ln M}\, \dv{\delta_c}\,,
\end{equation}
where \Nsub{} is the number of subhaloes within a given halo and $N_{\rm norm}$ is a normalization constant. The latter term, which is a free parameter that must be calibrated using simulations, corrects the total number count for progenitor subhaloes that exist at multiple redshifts. These structures are counted more than once in this approach \citep{schneider_structure_2015}. We calibrate $N_{\rm norm}$ at scales above the power spectrum cut-off to minimise any dependence on the shape of the power spectrum.

The form of the window function
can affect the predictions of the mean number of subhaloes in a halo of a given mass. Customarily, using a spherical top-hat function in real space has been favoured
because there is a simple relation between the filter volume, the filter scale radius, and the mass enclosed by the virialised object. Somewhat more importantly, this approach has also been shown to produce results that are in good agreement with those of numerical simulations of \CDM{} \citep{jenkins_mass_2001}.
However, for models of \DM{} with power spectra that are damped at small scales, such as the \MSM{}, this approach
predicts an excess of low-mass haloes compared to numerical simulations \citep{benson_dark_2013,schneider_structure_2015}.
This happens because the steep cut-off in the matter power spectrum permits only negligible contributions at high wavenumbers when calculating the variance of the matter density field in the limit of small radii. Meanwhile, in the same regime, the halo mass function is dominated by the derivative of the variance, which depends strongly on the form of the filter function that is used. Consequently, in \DM{} models with damped power spectra, the halo mass function calculated when adopting a real space top-hat filter function becomes independent of the power spectrum, and tends towards a constant value at small radii rather than becoming negligible.
Instead adopting a smooth filter in \textit{k}--space of the form,
\begin{equation}
    \tilde{W}\!\left(k \vert R\right) = \left[1+\left(kR\right)^{\hat{\beta}}\right]^{-1}\,,
\end{equation}
where \textit{k} is the wavenumber, \textit{R} is the filter scale radius, and $\hat{\beta} > 0$ is a free parameter,
results in better agreement with simulations \citep{leo_new_2018}.
To achieve this good agreement, the smooth \textit{k}--space filter introduces a second free parameter, $\hat{c}$, that connects the filter mass, \textit{M}, to \textit{R} through the mean density, $\bar{\rho}$, as $M\!\left(R\right) = 4 \pi \bar{\rho} \left(\hat{c}R\right)^3 \, /\, 3$. Both $\hat{\beta}$ and $\hat{c}$, and the normalization constant that we discussed earlier, $N_{\rm norm}$, must be calibrated with simulations. We showed in \citet{newton_constraints_2021} that using the \EPS{} formalism with a smooth \textit{k}--space filter calibrated using a subset of haloes from the \COCO{} suite \citep{bose_copernicus_2016,hellwing_copernicus_2016} produces results in good agreement with the simulations.

In this work we calibrate the \EPS{} and filter function parameters using gravity only counterparts of the `V2' and `V5' sets of Local Group-analogue simulations introduced by \citet{lovell_properties_2017}.
Each set comprises a \CDM{} volume and two \Ms[{\keV[7]}] \MSM{} volumes with $\sin^2\!\left(2\mixAng{}\right)\simeq 10^{-11}$. They were simulated using initial conditions that are identical to the first six simulation volumes of the \Apostle{} project \citep{fattahi_apostle_2016,sawala_apostle_2016}, except for the differences introduced by changing the properties of the \DM{}. All simulations adopt cosmological parameters from the \WMAP{} seventh-year data release \citep{komatsu_seven-year_2011}: \WMAPsevenp{}.

\begin{figure}
    \centering
    \includegraphics[width=\columnwidth]{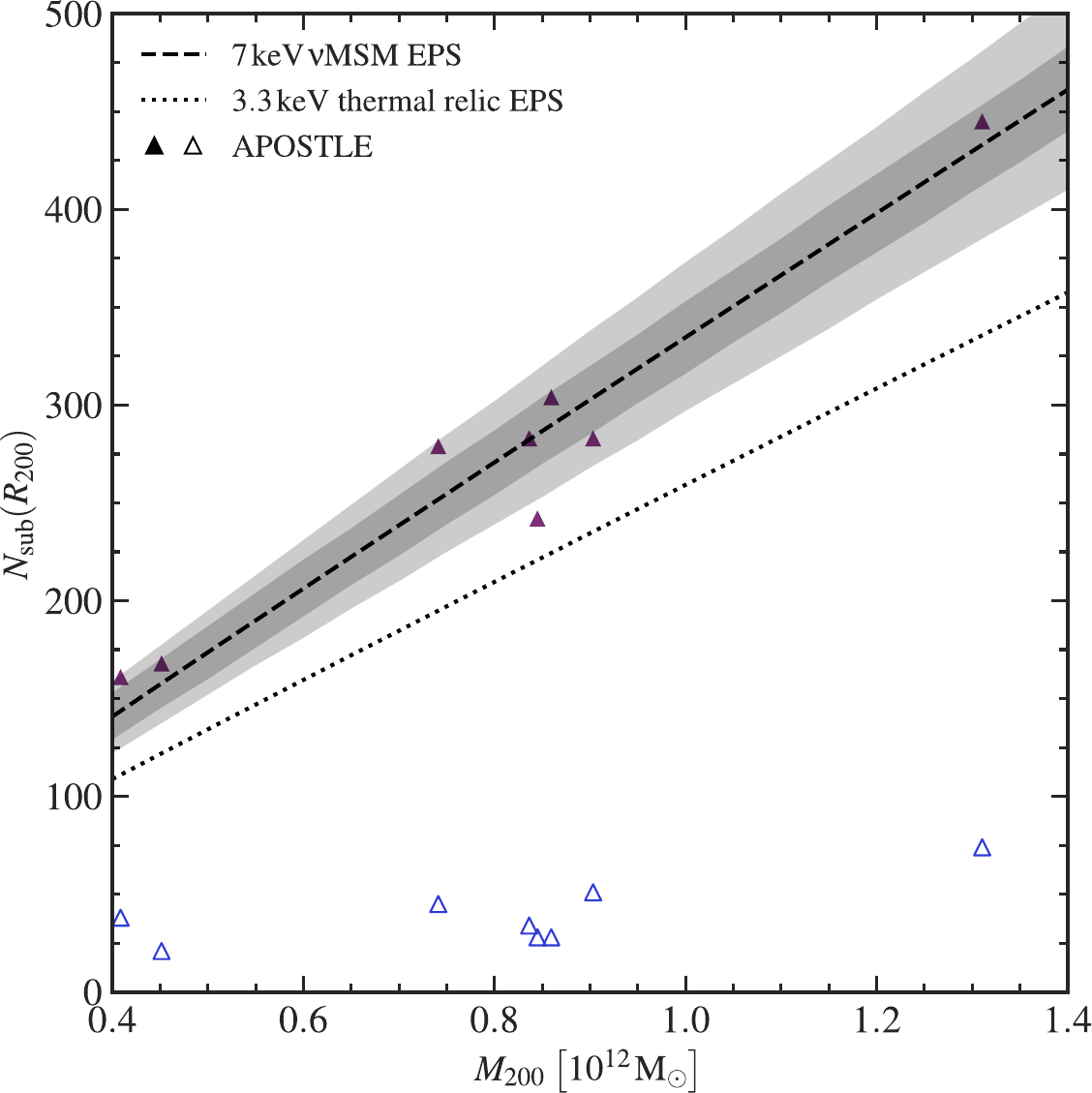}
    \caption{The total number of \DM{} subhaloes within \RNFW{} as a function of \DM{} halo mass, \MNFW{}. Need to state which nuMSM model parameters these are. The dashed line shows the mean number of subhaloes predicted by the \EPS{} formalism when adopting the \MSM{} parameters \Ms[7.1] and $\sin^2\!\left(2\mixAng{}\right)\simeq 10^{-11}$. The dark shaded region indicates the associated \percent{68} Poisson scatter, and the light shaded region gives the \percent{68} scatter modelled using a negative binomial distribution (see~\eqnrefp{eq:Methods:EPS:Neg_binom}). The symbols represent haloes from the \keV[7] \MSM{} \Apostle{} simulations for $\sin^2\!\left(2\mixAng{}\right)\simeq 10^{-11}$. Unfilled symbols are from subhalo catalogues where the prematurely destroyed subhaloes have not been recovered, and filled symbols show the same haloes after the prematurely destroyed subhaloes have been restored by following the procedure described in \secref{sec:Methods:Calibrate_EPS}. For comparison, we plot as a dotted line the number of subhaloes predicted by the \EPS{} formalism applied to the \keV[3.3] thermal relic \WDM{} model from \citet{newton_constraints_2021}.}
    \label{fig:Methods:Calibrate_EPS}
\end{figure}

Simulations of \DM{} models that impose a cut-off in the primordial power spectrum are susceptible to the formation of `spurious' haloes. These structures are produced by the artificial fragmentation of filaments caused by resolution-dependent gravitational instabilities that are generated by the discreteness of the simulation particles
\citep{wang_discreteness_2007,angulo_warm_2013,lovell_properties_2014}.
We identify and prune spurious haloes from the merger trees by adopting the procedure introduced by \citet{lovell_properties_2014}, which we summarize here.
Numerical gravitational instabilities emerge when the gravitational potential is sampled poorly by simulation particles. The onset of these instabilities is associated with a resolution-dependent mass threshold, and haloes that do not surpass this during their evolution are likely to be spurious.
In addition, spurious haloes typically form from highly aspherical Lagrangian regions in the initial conditions. These regions are characterized using the particles that compose the halo when it accumulates half of its maximum mass.
The shapes of the Lagrangian regions are parametrized by $s_{\rm half-max} = c/a$, where $c$ and $a$ are, respectively, the minor and major axes of the diagonalized moment of inertia tensor of the \DM{} particles in Lagrangian coordinates.
The threshold mass and $s_{\rm half-max}$ values for the
\MSM{} \Apostle{} are calculated in the appendix of \citet{lovell_properties_2017}, who found that almost all spurious haloes can be removed by requiring $M_{\rm max}{>}\Msunh[5.35\times10^7]$ and $\hSpher{}{>}0.165$. We adopt these thresholds to prune the \MSM{} \Apostle{} simulations we consider in \figref{fig:Methods:Calibrate_EPS}.

We also correct for other resolution-dependent effects. It has been shown that configuration-space structure finders fail to identify some small haloes that clearly survive to the present day. Subhaloes that are close to the centre of the host halo are particularly likely to be overlooked. Additionally, some small haloes are artificially disrupted by tidal forces due to numerical effects such as small particle number.
We recover all prematurely destroyed subhaloes by tracking their most-bound particles through the simulations and determining whether they survive to the present day. To do this we use the \citet{simha_modelling_2017} merging scheme implemented in the galaxy formation model \Galform{} \citep{cole_hierarchical_2000} that recovers the entire population of destroyed substructures, including objects that were disrupted by physical processes. We identify and remove the latter from the recovered population if they satisfy one of the following criteria:
\begin{itemize}
    \item The dynamical friction timescale, i.e. the time required for a subhalo to sink to the centre of the host halo, is less than or equal to the time elapsed since the subhalo fell below the resolution limit.
    \item The subhalo passes within the halo tidal disruption radius.
\end{itemize}
In both cases, the effects of tidal stripping and of interactions between other subhaloes are ignored. This procedure is discussed in more detail in \citet[\extapp{A}]{newton_constraints_2021}.

After pruning spurious haloes from the merger trees and recovering the prematurely destroyed subhaloes, we obtain excellent agreement between the \EPS{} predictions and the \MSM{} \Apostle{} \Nbody{} results using ${N_{\rm norm}=1.4},\, {\hat{\beta}=4.2},$ and ${\hat{c} = 3.9}$. These values are very similar to the values we found in \citet{newton_constraints_2021} when calibrating the \EPS{} formalism for thermal relic \WDM{}.
In \figref{fig:Methods:Calibrate_EPS} the dashed line shows the mean number of subhaloes predicted by the \EPS{} formalism within \RNFW{} of the centre of \MW{} mass \DM{} haloes in the \Ms[{\keV[7.1]}], $\sin^2\!\left(2\mixAng{}\right)\simeq 10^{-11}$ parameterizations of the \MSM{}.
Here and throughout the rest of this paper, \MNFW{} denotes the mass enclosed within \RNFW{}, the radius within which the mean density equals $200$ times the critical density.
The shaded regions represent the \percent{68} scatter in \Nsub{} at fixed halo mass modelled using a negative binomial distribution, which encapsulates the scatter seen in numerical simulations \citep{boylan-kolchin_theres_2010}.
This is given by
\begin{equation}
\label{eq:Methods:EPS:Neg_binom}
    {\rm P}\left(N \right|\left. r,\, p \right) =
    \frac{\upGamma\!\left(N+r\right)}
         {\upGamma\!\left(r\right)\upGamma\!\left(N+1\right)}\, p^r\! \left(1 - p\right)^N\,,
\end{equation}
where $\upGamma\!\left(x\right){=}\left(x-1\right)!$ and \textit{N} is the number of subhaloes. The parameter, ${p=\langle N \rangle\, /\, \upsigma^2,}$ where $\langle N\rangle$ and ${\upsigma^2=\sigmaPoisson^2 + \sigmaI^2}$ are, respectively, the mean and the dispersion of the distribution. The parameter, ${r=\sigmaPoisson^2\, /\, \sigmaI^2}$, describes the relative contribution of Poisson fluctuations compared to a second distribution that describes the intrinsic variability of the subhalo count within haloes of fixed mass. We model the scatter in the subhalo count by adopting $\sigmaI{} = 0.12 \langle N\rangle$, which is in good agreement with that found in \Nbody{} simulations such as \COCO{} \citep{newton_constraints_2021}.
In \figref{fig:Methods:Calibrate_EPS}, we also overlay the subhalo counts of \MW{} mass haloes in the \Apostle{} \Ms[{\keV[7]}], $\sin^2\!\left(2\mixAng{}\right)\simeq 10^{-11}$ simulations, both before and after restoring the population of prematurely destroyed subhaloes to the catalogues. The \EPS{} predictions and model of the scatter in the subhalo counts are in excellent agreement with the latter.

At fixed mass, the \MW{} mass haloes that form in the parameterization of the \MSM{} shown in \figref{fig:Methods:Calibrate_EPS} (dashed line) are predicted to host more \DM{} subhaloes than they would in the \keV[3.3] thermal relic \WDM{} model (dotted line). Consequently, larger fractions of the \MSM{} systems host at least as many subhaloes as the inferred total \MW{} satellite galaxy population. As \citet{newton_constraints_2021} were unable to place constraints on the \keV[3.3] thermal relic model using this approach, we will be similarly unable to constrain the $\sin^2\!\left(2\mixAng{}\right)\simeq 10^{-11}$ parameterizations of the \Ms[{\keV[7.1]}] \MSM{} model. We will discuss our full results in \secref{sec:Results:nuMSM_constraints:DM_only}.

\subsection{Calculating model acceptance probability}
\label{sec:Methods:Calculate_acceptance_probability}

We obtain our constraints on the \MSM{} parameter space by following the approach described in \citet{newton_constraints_2021}, which we summarize briefly here. First, we calculate the fraction, \fviable{}, of \MW{} mass systems that have at least as many satellites
\Nsat[\MSM{}],
as the total number of \MW{} satellite galaxies, \Nsat[MW], for a given set of \MSM{} parameters. This is given by
\begin{multline}
    \label{eq:Methods:f_v}
    \fviable{} = \int_0^\infty \dv{\Nsat[\rm MW]} p^{\rm MW}\!\left(\Nsat[\rm MW]\right) \times \\ \int_{\Nsat[\rm MW]}^{\infty} \dv{\Nsat[\MSM{}]}\, p^{\rm \MSM{}}\!\left(\Nsat[\MSM{}]\right)\,,
\end{multline}
where $p^{\rm MW}$ and $p^{\rm \MSM{}}$ are, respectively, the probability distributions of the \MW{} satellite count,
and of the number of
satellites
predicted to be hosted by \MW{} mass haloes at \z[0] in a given parameterization of the \MSM{} model.
Then, we obtain constraints by computing the intersection of the cumulative distribution of \fviable{} as a function of \MW{} halo mass at fixed \Ms{} and
\mixAng{}
with a five~per~cent rejection threshold. Parameterizations of the \MSM{} at and below this threshold are ruled out.

Our calculation of \fviable{} in \eqnref{eq:Methods:f_v} accounts for the scatter in the number of
\DM{} subhaloes
at fixed host halo mass and the uncertainty in the size of the total \MW{} satellite galaxy population.
As both of these quantities depend on the assumed mass of the Galactic halo, the viable fraction is also highly sensitive to it.
Estimates of the mass of the Galactic halo are still quite uncertain; indeed, it has been revised downwards in recent years \citep[see e.g.][]{bird_milky_2022}. To account for this uncertainty we calculate \fviable{} for several \MW{} halo masses spanning the most likely range determined from kinematic analyses of the Galactic potential. Unless stated otherwise the constraints we present in \secref{sec:Results} are marginalized over the distribution of \MW{} halo masses from \citet{callingham_mass_2019}.

\section{Constraints on the properties of sterile neutrinos in the \MSM{}}
\label{sec:Results}
The constraints on the parameter space of the \MSM{} depend on the assumed mass of the \MW{} halo, the rest mass of the sterile neutrino \DM{} candidate, \Ms{}, and the resonantly enhanced effective mixing angle of the sterile neutrino, \mixAng{}.
We consider \MW{} halo masses in the likely range
$\MMW{\Msun[{\left(0.5\text{--}2.0\right)\times10^{12}}]}$ \citep{callingham_mass_2019,cautun_milky_2020}, and values of $\sin^2\!\left(2\mixAng{}\right)$ between the \DM{} under- and over-production bounds if the sterile neutrino comprises all of the \DM{}. In this and subsequent sections we assume the \citet{planck_collaboration_planck_2020} cosmology: \PLANCKeighteenp{}.
For each combination of parameters, we compute the fraction of \MSM{} systems that produce at least as many
galaxies
as the number of \MW{} satellite galaxies inferred from observations using the methodology presented in \citet{newton_total_2018}, as implemented in \citet{newton_mw_2018}.

We obtain our most robust result, presented in \secref{sec:Results:nuMSM_constraints:DM_only}, under the premise that all \DM{} subhaloes host a visible galaxy; therefore, we make no assumptions at all about galaxy formation processes. This maximises \fviable{} and produces extremely robust lower limits on the viable \MSM{} parameter space.
In \secref{sec:Results:nuMSM_constraints:Galform}, we use the \Galform{} semi-analytic model of galaxy formation to estimate the number of satellite galaxies
in each parameterization of the \MSM{}. This allows us to account for astrophysically relevant processes, such as reionization, that can affect the properties of the low-mass galaxy population.
In \secref{sec:Results:nuMSM_constraints:Num_MW_sats}, we consider how our results are affected when we use different estimates of \Nsat[MW] from that reported by \citet{newton_total_2018} and compare these results with others in recent literature.

\subsection{Constraints from structure formation considerations}
\label{sec:Results:nuMSM_constraints:DM_only}

\begin{figure}
	\includegraphics[width=\columnwidth]{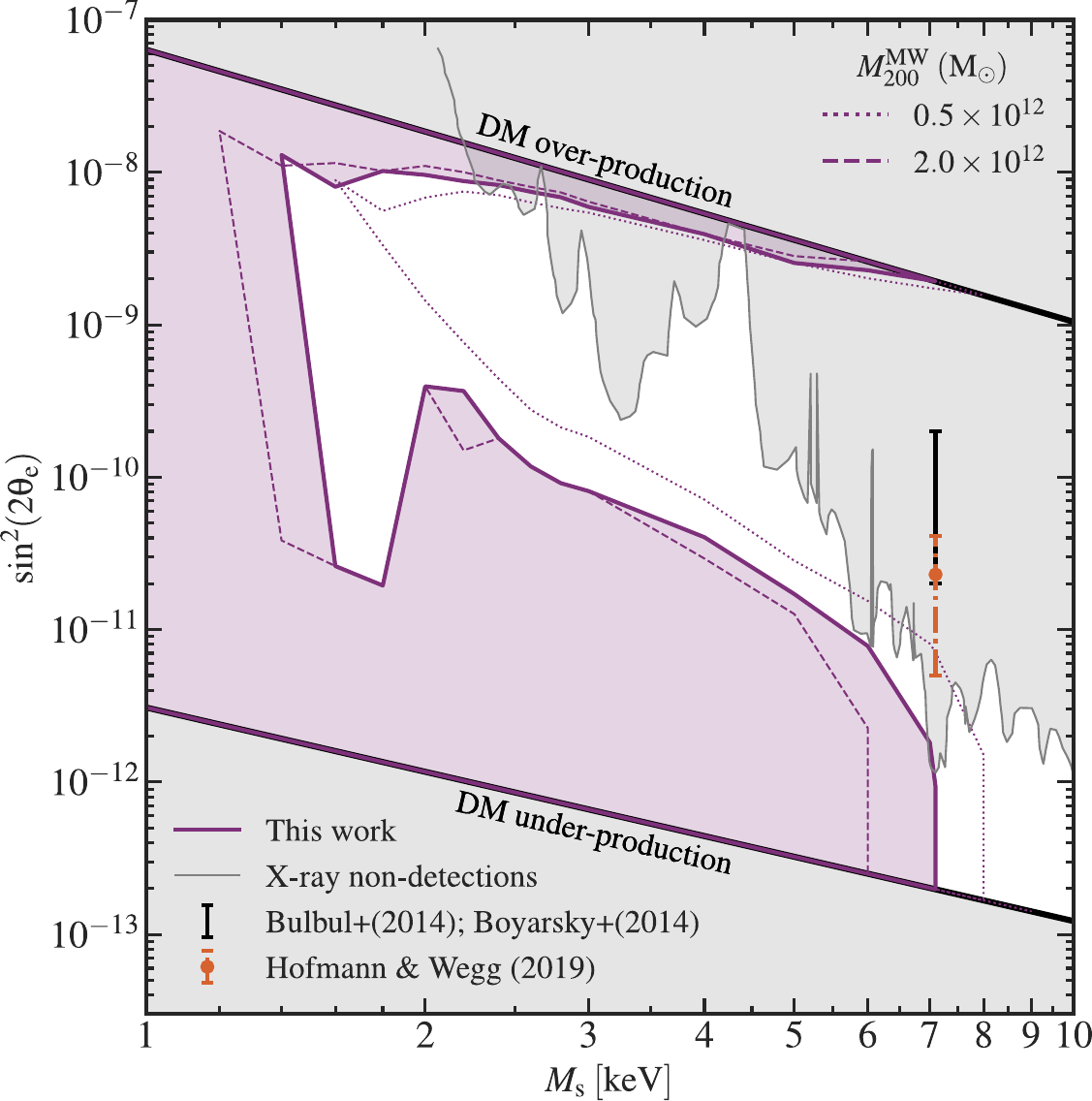}%
	\caption{%
    Constraints on the \MSM{} parameter space obtained assuming all subhaloes host a galaxy. Our fiducial structure count result (thick solid curve) is marginalized over the uncertainties in the \citet{callingham_mass_2019} estimate of the \MW{} halo mass. Parameter combinations in the shaded region to the left of this envelope are ruled out with at least \percent{95} confidence. We also show the constraint envelopes we obtain assuming \MW{} halo masses of \MMW{\Msun[{0.5\times10^{12}}]} (dotted curve) and \MMW{\Msun[{2.0\times10^{12}}]} (dashed curve).
     The shaded regions in the upper and lower portions of the parameter space delimit parameter combinations for which the \DM{} is over- or under-produced compared with current observational estimates of the composition of the Universe if resonantly produced sterile neutrinos compose all of the \DM{} \citep{asaka_msm_2005,schneider_astrophysical_2016}.
     The lower limits on the effective mixing angle obtained from X-ray non-detections are shown using the thin solid curve and shading \citep{neronov_decaying_2016,ng_new_2019,roach_nustar_2020,roach_long-exposure_2023}.
     The solid error bar indicates the parameter range favoured by X-ray flux measurements if the \DM{} consists of \Ms[{\keV[7.1]}] decaying sterile neutrinos \citep{boyarsky_unidentified_2014,bulbul_detection_2014}.
     The dash--dot error bar shows the preferred range determined from X-ray line detections in the Galactic bulge \citep{hofmann_71_2019}.
     }
     \label{fig:Results:constraints_eps_marginalized}
\end{figure}

In \figref{fig:Results:constraints_eps_marginalized}, we present constraints on the parameter space of \MSM{} models obtained by comparing the abundance of \DM{} substructure predicted using the \EPS{} formalism with the \citet{newton_total_2018} number count of \MW{} satellite galaxies. Parameter combinations of \Ms{} and \mixAng{} within the shaded regions are ruled out with at least \percent{95} confidence when marginalizing over the uncertainties in the \MW{} halo mass from \citet{callingham_mass_2019}. 
We show recent astrophysical limits from
X-ray modelling of M31 \citep{horiuchi_sterile_2014}, Galactic bulge non-detections \citep{roach_nustar_2020}, and blank sky non-detections \citep{dessert_dark_2020,foster_deep_2021,roach_long-exposure_2023} using a dark grey curve and shaded region.
The viable parameter space decreases significantly at and below \Ms[{\keV[7]}], and we rule out all parameterizations of the \MSM{} with \MsConstraint{1.2}. This result is valid for all \MW{} halo masses we consider up to and including the largest mass of \MMW{\Msun[{2\times10^{12}}]} (dashed curve), which produces the least constraining results.
When marginalizing over the uncertainty in the \citet{callingham_mass_2019} estimate of the \MW{} halo mass, we rule out all parameterizations of the \MSM{} with \MsConstraint{1.4}.
These results are consistent with constraints obtained by theoretical and observational analyses of the phase space density of \DM{} within \MW{} dwarf satellite galaxies \citep{boyarsky_lower_2009}.

As we discussed in \secref{sec:Introduction}, the \keV[3.55] excess observed emanating from several \DM{}-rich astrophysical systems could potentially be evidence of decaying \MSM{} sterile neutrinos with rest masses, \Ms[{\keV[7.1]}].
If all of the \DM{} in the Universe is composed of these \MSM{} neutrinos, X-ray flux observations favour a narrow window of viable parameter space \citep{boyarsky_unidentified_2014,boyarsky_checking_2015, bulbul_detection_2014,iakubovskyi_testing_2015,ruchayskiy_searching_2016}. We indicate this in \figref{fig:Results:constraints_eps_marginalized} using a solid error bar.
We also show the preferred region of parameter space determined from X-ray line detections in the Galactic bulge \citep{hofmann_71_2019} using a dash--dot error bar.
Our fiducial results, which we emphasize make no assumptions about galaxy formation physics, constrain the \Ms[{\keV[7.1]}] \MSM{} parameter space only at small values of \mixAng{} because the amount of substructure produced by other parameterizations is too great.
If the mass of the \MW{} is at the lower end of estimates (dotted curve), as suggested by recent studies \citep[e.g.][]{bird_milky_2022,necib_substructure_2022-1,zhou_circular_2023,roche_escape_2024}, more of the parameter space is excluded, including some parameterizations favoured by \citet{hofmann_71_2019}.
The limits from X-ray non-detections already exclude the \MSM{} as a viable explanation for the origin of the \keV[3.55] excess, although they are subject to modelling uncertainty \citep[][though see \citealp{abazajian_technical_2020,boyarsky_technical_2020}, and \citealp{dessert_response_2020}]{dessert_dark_2020}. Combining this exclusion region with our fiducial satellite count constraints, all parameterizations of the \MSM{} with \Ms[{\keV[7.1]}] are excluded.
This conservative and highly robust result can be extended to more of the parameter space by adopting a physically motivated prescription to populate some of the subhaloes with galaxies. This reduces the subhalo occupation fraction and, consequently, strengthens the constraints on the viable parameter space of favoured models.
We explore this in \secref{sec:Results:nuMSM_constraints:Galform}.

Our analysis depends on the total number of \MW{} satellite galaxies, which is uncertain. In \appref{sec:Appendix:Structure_formation_constraints_N_sats}, we present the structure count constraints we obtain when assuming different values of the size of the total \MW{} satellite galaxy population that have been proposed in recent years. We will discuss this within the context of applying a galaxy formation model in \secref{sec:Results:nuMSM_constraints:Num_MW_sats}.

\subsection{Modelling galaxy formation processes}
\label{sec:Results:nuMSM_constraints:Galform}
The satellite count constraints presented in \secref{sec:Results:nuMSM_constraints:DM_only} are highly robust and conservative lower limits on the \MSM{} parameter space. However, they ignore the effects of baryon physics on the formation and evolution of the stellar component in low-mass haloes. Although the details of these processes are not fully understood they are expected to have a significant effect on the properties of the satellite galaxy population of the \MW{}.
To understand how baryonic physics affects the constraints we explore the parameter space of these processes using the \Galform{} semi-analytic model of galaxy formation \citep{cole_recipe_1994,cole_hierarchical_2000,lacey_unified_2016}. The parameters of this state-of-the-art model are calibrated to reproduce a variety of observational results, including the \z[0] luminosity functions of galaxies in the local Universe, the \ion{H}{I} mass function, the distributions of galaxy morphologies and sizes, the \z[0] Tully--Fisher relation, and the number counts of sub-millimeter galaxies. The complete set of observations used to calibrate the \Galform{} model parameters is described in \citet[][\extsec{4.2}]{lacey_unified_2016}.
Although \Galform{} has been calibrated assuming the \LCDM{} model, it is not necessary to recalibrate it for each parameterization of the \MSM{}. This is because \Galform{} is calibrated using massive galaxies far above the mass scales relevant here, and we do not expect that changes in the small-scale dark matter distribution will affect the outcome.

Among the various astrophysical processes that can be explored with \Galform{}, the effect of the reionization of the Universe on low-mass galaxies is of particular interest for this study. During reionization the intergalactic medium is heated by the UV background radiation and is prevented from cooling into low-mass haloes. This inhibits the accretion of fresh supplies of cold gas that are necessary for star formation, thereby curtailing further growth of the stellar component.
In \Galform{} this is modelled by assuming that after reionization finishes at redshift, $\z[{\zreion{}}]$, gas cooling does not take place in haloes with circular velocities, $\Vcirc{} < \Vcut{}$. In general, more low-luminosity galaxies are able to form if reionization finishes later, and lower values of the circular velocity cooling threshold, \Vcut{}, permit those galaxies to become brighter because they are able to continue accreting gas for longer \citep[e.g.][]{bose_imprint_2018}. Consequently, we expect to find the most stringent constraints on the \MSM{} in the parameterizations of \Galform{} where reionization finishes at higher redshifts and \Vcut{} is large. Following the approach of \citet{newton_constraints_2021}, we explore the parameter space of reionization spanning all combinations of $\zreion[{\left[6,7,8\right]}]$ and $\Vcut[{\kms[{\left[25,30,35\right]}]}]$.

\begin{figure}%
    \centering%
	\includegraphics[width=\columnwidth]{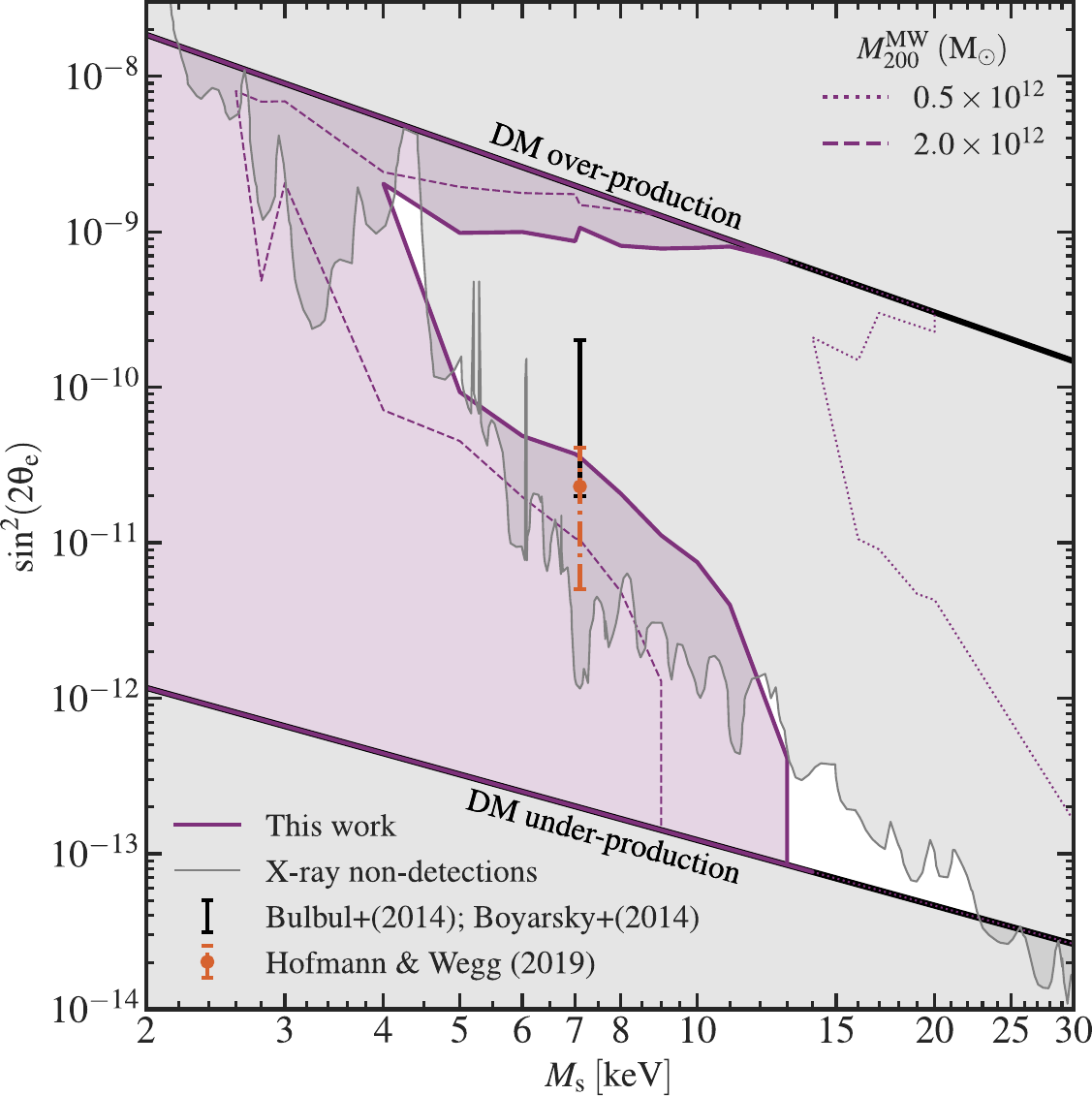}%
	\caption{Constraints on the \MSM{} parameter space obtained assuming our fiducial model of reionization (\zreion[7] and \Vcut[{\kms[30]}]) in the \Galform{} galaxy formation model. Our main result marginalizes over the uncertainties in the \citet{callingham_mass_2019} estimate of the \MW{} halo mass (thick solid curve). Parameter combinations in the shaded region to the left of this envelope are ruled out with at least \percent{95} confidence. We also show the constraint envelopes we obtain assuming \MW{} halo masses of \MMW{\Msun[{0.5\times10^{12}}]} (dotted curve) and \MMW{\Msun[{2.0\times10^{12}}]} (dashed curve).
    As in \figref{fig:Results:constraints_eps_marginalized}, the parameter space is delimited by upper and lower shaded regions in which the \DM{} is over- or under-produced compared with current observational estimates of the composition of the Universe \citep{asaka_msm_2005,schneider_astrophysical_2016}.
     Lower limits on the effective mixing angle obtained from X-ray non-detections are shown using the thin solid curve and shading \citep{neronov_decaying_2016,ng_new_2019,roach_nustar_2020,roach_long-exposure_2023}. The solid and dot--dashed error bars indicate the parameter ranges favoured by X-ray flux measurements from different systems if all of the \DM{} comprises \Ms[{\keV[7.1]}] sterile neutrinos \citep{boyarsky_unidentified_2014,bulbul_detection_2014,hofmann_71_2019}.}%
 \label{fig:Results:constraints_galform_fiducial_xray}%
\end{figure}%

\begin{figure*}
    \centering%
	\includegraphics[width=\textwidth]{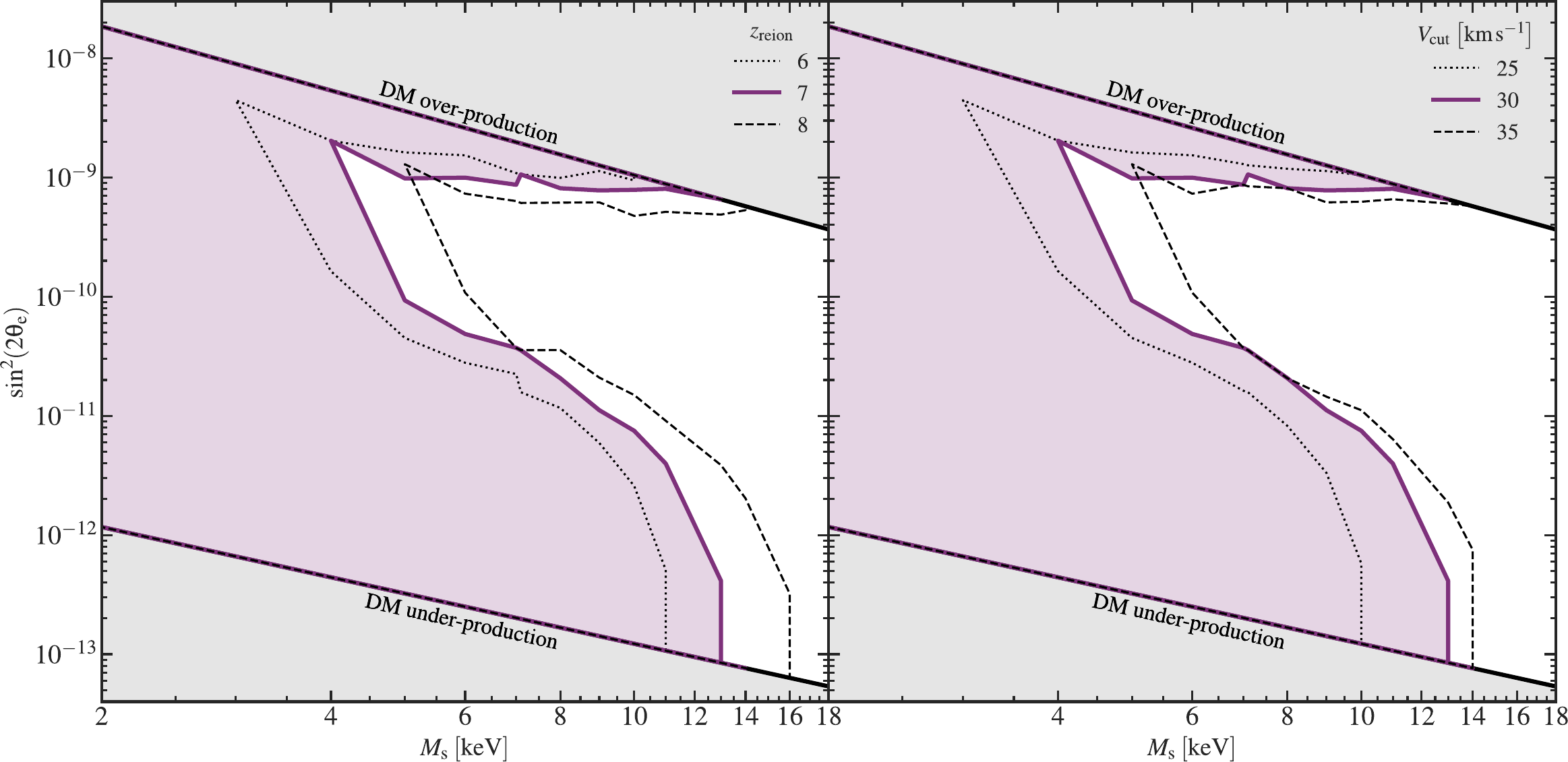}%
	\caption{Constraints on the \MSM{} parameter space obtained assuming different parameterizations of reionization in the \Galform{} galaxy formation model when marginalizing over the uncertainties in estimates of the \MW{} halo mass from \citet{callingham_mass_2019}. Parameter combinations to the left of the envelopes are ruled out with at least \percent{95} confidence. In both panels our fiducial result is indicated by the thick solid curve and shaded region.
    \textit{Left panel}: The circular velocity cooling threshold is fixed at \Vcut[{\kms[30]}] and we compute constraint envelopes assuming reionization finishes at \zreion[6] (dotted curve), \zreion[7] (solid curve), and \zreion[8] (dashed curve).
    Parameterizations in which reionization finishes earlier produce more stringent constraints on the \MSM{} parameter space.
    \textit{Right panel}: The redshift at which reionization finishes is fixed at \zreion[7] and we compute constraint envelopes assuming circular velocity cooling thresholds of \Vcut[{\kms[25]}] (dotted curve),
    \kms[30] (solid curve), and
    \kms[35] (dashed curve).
    Higher cooling thresholds produce more stringent constraints on the \MSM{} parameter space.}%
	\label{fig:Results:constraints_galform_explore_reion}%
\end{figure*}%

To obtain constraints on the \MSM{}, we generate predictions of the number of satellite galaxies brighter than \MV[0] in \MW{} mass haloes and compare them with estimates of the total \MW{} satellite complement based on observations \citep{newton_total_2018}. We adopt
the \citet{lacey_unified_2016} version of the \Galform{} model, which
provides a reasonable description of a large variety of baryonic feedback and evolutionary processes, including those that are particularly relevant for low-mass galaxies such as star formation; gas heating and cooling; supernovae feedback; and the chemical enrichment of stars and gas; and vary the reionization parameters as described above.
We generate Monte Carlo merger trees using an improved version of the \citet{parkinson_generating_2008} algorithm,
which we calibrate to match as closely as possible the merger trees obtained from
recent \CDM{} \Nbody{} simulations (A.~J.~Benson et al., in preparation).
In \citet{newton_constraints_2021} we noted that the match to the \Nbody{} results at the faint end is inexact because the algorithm lacks sufficient free parameters to model both the high- and low-mass ends of the mass function. Here, we try to address this by modifying the \citet{parkinson_generating_2008} algorithm to include an explicit dependence on the linear growth-rate factor and the logarithmic slope of the \SigmaFunc{M} relation, which describes the variance of the matter density field as a function of mass. This introduces three new free parameters that we calibrate using \Nbody{} simulations. We provide full details of our modifications to the merger tree algorithm in \appref{sec:Appendix:Modifications_to_PCH}. Our alterations significantly improve the agreement between the Monte Carlo satellite galaxy luminosity functions and the \Nbody{} results; however, they are still slightly discrepant. Therefore, we adopt the approach described in \citet{newton_constraints_2021} and introduce an empirical correction to remap the \MV{} values of the Monte Carlo satellite galaxies to new values so that the resulting luminosity functions are consistent with those from the \Nbody{} simulations.
Although we cannot eliminate the need to apply an empirical correction,
the modifications we make to the \citet{parkinson_generating_2008} algorithm reduce the size of the correction we must introduce when compared with that required in \citet{newton_constraints_2021}. These are typically \percent{10}.

To generate the constraints on the \MSM{} we produce
satellite galaxy luminosity functions for $500$ realizations of each \MW{} halo mass that we consider and use them to calculate the model acceptance distributions in the manner we described in \secref{sec:Methods:Calculate_acceptance_probability}. Parameterizations of the model for which no more than \percent{5} of the realizations have at least as many satellite galaxies as the observations are ruled out. In \figref{fig:Results:constraints_galform_fiducial_xray}, the shaded exclusion region shows the constraints we obtain on the \MSM{} parameter space
using our fiducial model of reionization with \zreion[7] and \Vcut[{\kms[30]}]
when marginalizing over the uncertainties in the \MW{} halo mass from \citet{callingham_mass_2019}. Parameter combinations in this region are ruled out with at least \percent{95} confidence.
Once again, we show the constraint envelopes for our fiducial parameterization of reionization assuming \MW{} halo masses at the lower (dotted curve) and upper (dashed curve) ends of the likely range.
Independently of \MW{} halo mass, all parameterizations of the \MSM{} with \MsConstraint{2.6} are excluded by the \MW{} satellite count constraints with at least \percent{95} confidence (dashed curve).
When marginalizing over possible \MW{} halo masses, almost all of the \Ms[{\keV[7.1]}] \MSM{} parameter space favoured by \citet{hofmann_71_2019} as an explanation for the \keV[3.55] line, and some of the parameter space favoured by \citet{bulbul_detection_2014} and \citet{boyarsky_unidentified_2014}, is excluded.
These constraints complement the reported X-ray non-detection lower limits in closing the \Ms[{\keV[7.1]}] \MSM{} parameter space.
If the \MW{} mass is \MMW{\Msun[{0.5\times10^{12}}]}, we rule out all parameterizations of the \MSM{} with \MsConstraint{14}.
All parameterizations of the \MSM{} with \Ms[{\keV[7.1]}] are ruled out when $\MMW{}\leq\Msun[{8\times10^{11}}]$ (not shown).

We explore how changing the parameterization of reionization affects the constraints in \figref{fig:Results:constraints_galform_explore_reion}. The left panel shows the effect of varying the redshift at which reionization finishes while holding the circular velocity cooling threshold fixed at \Vcut[{\kms[30]}]. A late-finishing epoch of reionization (dotted curve) enables more galaxies to form in low-mass \DM{} haloes and weakens the constraints on the \MSM{} parameter space. This is most pronounced in parameterizations of the \MSM{} with $\Ms{}\geq\keV[10]$, where the parameter space is almost completely unconstrained.
In the scenario in which reionization finishes earlier (dashed curve) than our fiducial choice the viable parameter space closes slightly, and all parameterizations below \Ms[{\keV[5]}] are excluded.
The right panel shows the constraints for scenarios in which reionization finishes at \zreion[7]. Lowering the circular velocity cooling threshold permits a larger fraction of the low-mass galaxy population to accrete fresh supplies of cold gas after the end of reionization. This is then available to fuel further star formation that causes the faintest galaxies to become brighter than \MV[0] by \z[0]. The brightened galaxies populate the satellite galaxy luminosity function of the halo and weaken the constraints on the \MSM{} parameter space (dotted curve). Raising \Vcut{} has the opposite effect, and a larger fraction of the low-mass galaxies may only form stars from the reservoir of cold gas accreted prior to reionization. This limits how bright they become by \z[0], and fewer of them become brighter than \MV[0] to populate the luminosity function. Consequently the constraints become stronger in this case (dashed curve).
Similarly to the left panel, parameterizations of the \MSM{} with sterile neutrino rest masses larger than \Ms[{\keV[10]}] are particularly sensitive to changes to the circular velocity cooling threshold.

\subsection{Dependence on the total number of Milky Way satellite galaxies}
\label{sec:Results:nuMSM_constraints:Num_MW_sats}

\begin{figure}%
    \centering%
	\includegraphics[width=\columnwidth]{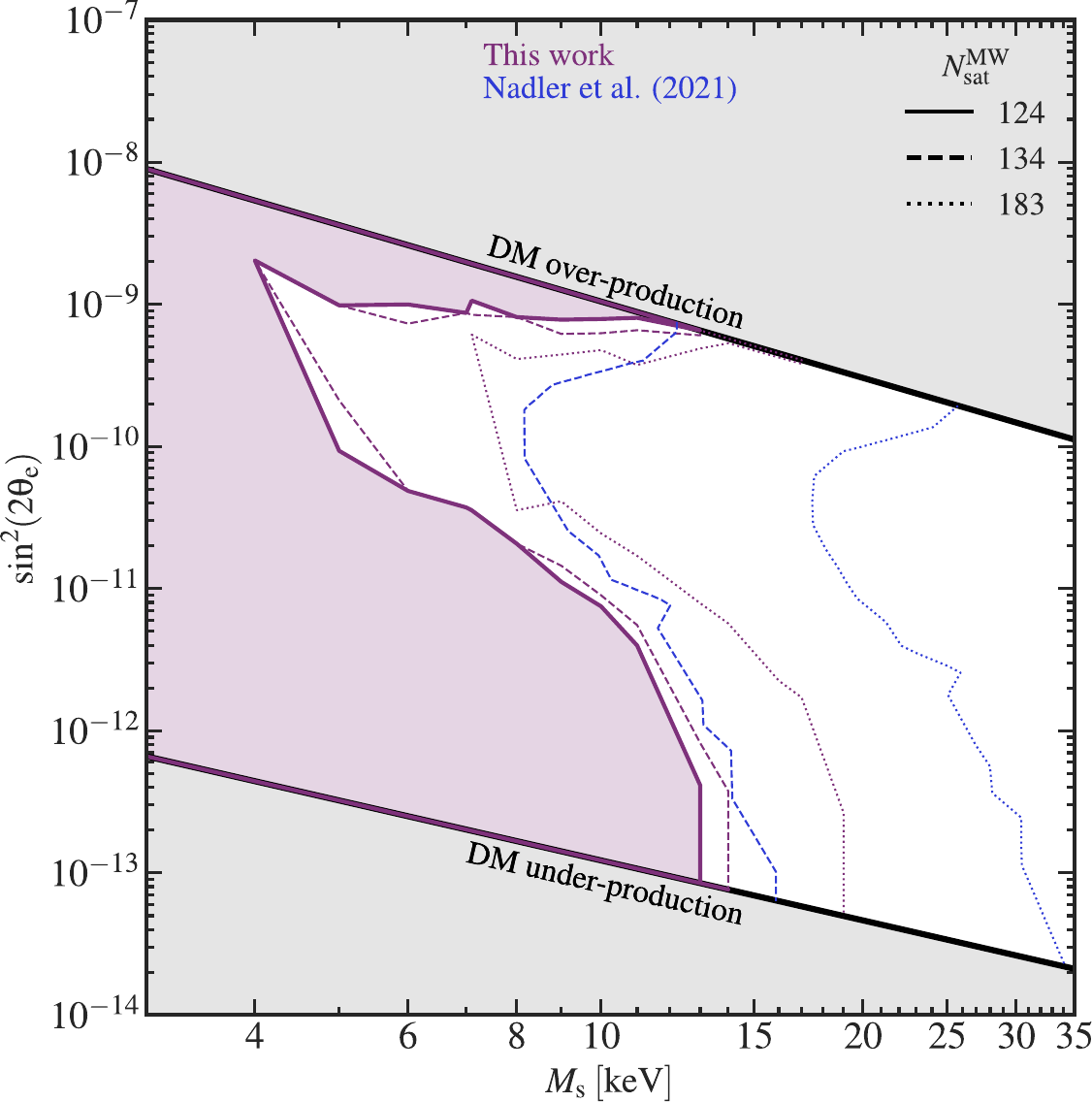}%
	\caption{Constraints on the \MSM{} parameter space from analyses of \MW{} satellite counts. The shaded regions in the upper and lower portions of the parameter space delimit parameter combinations for which the \DM{} is over- or under-produced compared with current observational estimates of the composition of the Universe if resonantly produced sterile neutrinos compose all of the \DM{} \citep{asaka_msm_2005,schneider_astrophysical_2016}. Our results adopt our fiducial model of reionization (\zreion[7] and \Vcut[{\kms[30]}]) in the \Galform{} galaxy formation model marginalized over the uncertainties in estimates of the \MW{} halo mass from \citet{callingham_mass_2019}. We compute constraint envelopes assuming a total \MW{} satellite galaxy population with a mean value of \NsatNum[MW]{124} (thick solid curve), $134$ (dashed curve), $183$ (dotted curve), and $270$ (dot-dashed curve). Parameter combinations within the envelopes are ruled out with at least \percent{95} confidence. For comparison we also show the constraint envelopes from \citet{nadler_constraints_2021}.}%
	\label{fig:Results:constraints_galform_varying_NsatMW}%
	\vspace{-10pt}%
\end{figure}

Recent improvements in the quantity and quality of observations of the \MW{} satellite system have motivated renewed efforts to estimate the size of the total population, \Nsat[MW].
This number remains highly uncertain because current observations are insufficiently deep to probe as far as the virial radius of the \MW{} halo, and they do not cover the entire sky.
As analyses such as ours depend on
\Nsat[MW]
it is important to understand how the results are affected when assuming different values. This is especially important when comparing our results with other studies that implement
the same technique to impose constraints on the parameter space of the \MSM{}
\citep{nadler_constraints_2021}. In \figref{fig:Results:constraints_galform_varying_NsatMW}, we present the constraints derived from our fiducial \Galform{} model (\zreion[7] and \Vcut[{\kms[30]}]) while adopting three different estimates of the size of the \MW{} satellite galaxy population from recent literature \citep{newton_total_2018,nadler_modeling_2019,nadler_milky_2020}.
As in previous sections we have marginalized over the uncertainties in the \citet{callingham_mass_2019} estimate of the \MW{} halo mass.
Our constraints become more restrictive as the assumed value of \Nsat[MW]
increases
because the \MSM{} models must produce more substructure containing galaxies to remain viable. Adopting \NsatNum[MW]{134} (dashed curve) as suggested by \citet{nadler_modeling_2019}, a value that is slightly larger than the \citet{newton_total_2018} estimate we used for our fiducial results in \secrefs{sec:Results:nuMSM_constraints:DM_only}{sec:Results:nuMSM_constraints:Galform}, yields little change in the constraints for parameterizations of the \MSM{} with \Ms[{\keV[7.1]}].
If the total satellite population is closer to \NsatNum[MW]{183} (dotted curve), as suggested by \citet{nadler_milky_2020},
we find that the viable parameter space of the
\MSM{} decreases significantly, in agreement with \citet{nadler_constraints_2021}. Moreover, the currently favoured parameter space of plausible \MSM{} models corresponding to the putative \keV[3.55] X-ray excess closes completely.

Alongside our constraints in \figref{fig:Results:constraints_galform_varying_NsatMW}, we plot the constraint envelopes reported by \citet{nadler_constraints_2021}, who adopt the estimates of the size of the \MW{} satellite galaxy population mentioned above. At fixed \Nsat[MW]{} their constraints are more stringent than ours. A comparison between our results is complicated by various factors. To begin with, \citet{nadler_constraints_2021} do not calculate constraints on the \MSM{} directly but rather infer them indirectly from constraints on the mass of a thermal relic neutrino.

There are several other differences between the two analyses. \citet{nadler_constraints_2021} adopt the \citet{venumadhav_sterile_2016} computation of the momentum distributions, which results in systematically larger cut-offs in the primordial power spectra at fixed sterile neutrino parameters than the \citet{laine_sterile_2008} algorithm that we use \citep{lovell_anticipating_2023}.
\citeauthor{nadler_constraints_2021} make use of subhalo abundance matching and several analytic prescriptions to populate a subset of subhaloes with galaxies in simulations of {\it thermal relic} neutrinos. They compare this with their estimate of \Nsat[MW], which they determine using a forward modelling framework that accounts for the tidal influence of the \LMC{} on the \MW{} halo, to obtain a constraint on the thermal relic \WDM{} parameter space. 
They  express this result in terms of the half-mode mass, \Mhm{}, which is the mass corresponding to the half-mode scale, $k_{\rm hm}$, at which
the thermal relic power spectrum, $P_{\rm WDM}$, is suppressed by a factor of four relative to the \CDM{} power spectrum, $P_{\rm CDM}$ \citep{viel_warm_2013,bose_copernicus_2016}, in other words, the scale at which 
\begin{equation}
    \sqrt{\frac{P_{\rm WDM}\!\left(k_{\rm hm}\right)}{P_{\rm CDM}\!\left(k_{\rm hm}\right)}} = 0.5\,.
\end{equation}
\citet{nadler_constraints_2021} remap their thermal relic constraint to the \MSM{} parameter space by comparing \Mhm{} computed for each parameterization of the \MSM{} to that of the coldest permitted thermal relic power spectrum, and rule out those parameterizations where \Mhm{} is greater than that of the thermal relic model.

The indirect  approach followed by \citet{nadler_constraints_2021}  is not directly comparable with our method. However, the stringency of their constraints suggests that their combination of subhalo abundance matching and analytic prescriptions
suppresses the formation of low-mass galaxies more aggressively than the \Galform{} prescription we adopt. As we discuss in detail in \secref{sec:Discussion}, this may be connected to the fixed mass threshold \citeauthor{nadler_constraints_2021} employ to determine whether a galaxy is able to form in a given (sub)halo.
Somewhat surprisingly, even when we adopt extremely high values of \NsatNum[MW]{} we are unable to replicate the stringency of their constraints obtained using smaller values.
Some of this discrepancy could be explained by our choice to use the \citet{newton_total_2018} probability distribution to describe the uncertainty in \NsatNum[MW]{}, which is broader than that of \citet{nadler_milky_2020} on which the constraints reported in \citet{nadler_constraints_2021} are based.
Broader uncertainties on \Nsat[MW]
can weaken the constraints by up to \percent{15} \citep[see][\extsec{2.4}]{newton_constraints_2021}. However, the size of the discrepancy between the results of the two analyses suggests that this is a subdominant effect.

While a detailed investigation of the sources of disparity between the two studies is beyond the scope of this work,
as we will see in \secref{sec:Discussion},
the methodological disparities in populating subhaloes with galaxies likely play a significant role alongside uncertainties in the particle physics calculation.
The results we present here
support the view that our approach produces the most conservative constraints on the parameter spaces of alternative \DM{} models relative to the standard \LCDM{} paradigm.
Efforts such as these complement
various observation-based analyses such as measurements of the X-ray spectra of \DM{}-dominated systems, the \Lyman{\alpha} forest \citep{irsic_unveiling_2024}, and gravitational lensing \citep{enzi_joint_2021,zelko_constraints_2022}.

\section{Discussion}
\label{sec:Discussion}
The \MSM{} is an appealing, physically motivated alternative to the \LCDM{} paradigm. It provides a plausible explanation for the extremely small masses of the active neutrinos in the Standard Model of particle physics and introduces a mechanism to produce the matter--anti-matter asymmetry in the Universe. Through these processes it can produce a \DM{} particle candidate in a plausible range of parameter space that appears to be consistent with astrophysical constraints. In certain parameterizations of the \MSM{} the sterile neutrino \DM{} behaves relativistically at early times and suppresses the formation of \DM{} structure at and below the scale of dwarf galaxies. Such departures from the predictions of \LCDM{} could manifest as a suppression of the number of low-mass satellite galaxies around \MW{} mass hosts.
This provides a useful metric to evaluate the viability of different parameterizations of the model \citep[this approach is described in detail in][]{newton_constraints_2021}. We exploit this to constrain the two-dimensional parameter space of the \MSM{}.

The choices that are adopted to populate \DM{} substructure with galaxies play a particularly important role in setting the stringency of the constraints.
Scenarios in which the subhalo occupation fraction is high exclude less of the model parameter space because more parameterizations are consistent with the inferred total number of \MW{} satellite galaxies.
We rule out a large fraction of the parameter space below \Ms[{\keV[7.1]}] when assuming that all \DM{} substructure hosts galaxies (see \figref{fig:Results:constraints_eps_marginalized}); an extreme assumption that is likely to be physically unrealistic. The advantage of this choice is that the results provide robust lower limits that are independent of assumptions about the details of galaxy formation physics.
The rest of the parameter space is largely unconstrained; in this regime the most compelling constraints come from observational analyses of X-ray non-detections, although these are subject to modelling uncertainty \citep{horiuchi_sterile_2014,neronov_decaying_2016,ng_new_2019,dessert_dark_2020,roach_nustar_2020,foster_deep_2021,roach_long-exposure_2023}.

Adopting physically motivated galaxy formation prescriptions, as we do to obtain the results in \figref{fig:Results:constraints_galform_fiducial_xray}, lowers the subhalo occupation fraction and rules out large swathes of parameter space below \Ms[{\keV[10]}], and all parameterizations at and below \Ms[{\keV[4]}]. In both cases, our fiducial results suggest that \MW{} satellite counts alone are insufficient to exclude all parameterizations of the \MSM{} with \Ms[{\keV[7.1]}], which is favoured as an explanation for the \keV[3.55] excess detected in the X-ray spectra of some galaxies and clusters.
In both scenarios, the shapes of the exclusion regions are affected by the non-monotonic evolution of the momentum distribution of the \DM{} as a function of \mixAng{} at fixed \Ms{}. The coldest momentum distributions are typically found at the logarithmic midpoint between the \DM{} under- and over-production bounds.

The constraints could be strengthened further if the mass of the \MW{} halo is lower than currently assumed. We determine our fiducial results by marginalizing over the uncertainties in the \citet{callingham_mass_2019} estimate of \MMW{\Msun[{1.17^{+0.21}_{-0.15}\times10^{12}}]}. The closest recent comparable analysis to ours assumes \MMW{\Msun[{1.4\times10^{12}}]} \citep{nadler_constraints_2021}. A synthesis of
the mass estimates spanning the past two decades
favours \MMW{\Msun[{8.5\times10^{11}}]} \citep{bird_milky_2022}, although this is \percent{50} more massive than the results derived from recent analyses using the extensive \Gaia{} data set of six-dimensional phase space information about the \MW{} stellar halo. Computing our constraints using \MMW{\Msun[{0.5\times10^{12}}]} improves the stringency of our results, which is especially
dramatic when we adopt the \Galform{} prescription to populate the subhaloes with galaxies. In this case, we rule out all parameterizations of the \MSM{} at and below \Ms[{\keV[14]}] with at least \percent{95} confidence and strongly exclude \MSM{} sterile neutrinos as the sole explanation for the \keV[3.55] excess. The latter finding holds for nearly all parameterizations of reionization except in the late-finishing scenarios with \zreion[6] and $\Vcut[{\kms[25]}]$ (not shown). A recent analysis of the \Lyman{\beta} forest suggests that reionization may finish as late as \zreion[5.5] \citep{zhu_long_2022}. Our results (see \figref{fig:Results:constraints_galform_explore_reion}) suggest that the \Ms[{\keV[7.1]}] parameter space is unlikely to be fully closed by \MW{} satellite counts in this case; however, in combination with the lower limits from X-ray non-detections (also shown in \figrefs{fig:Results:constraints_eps_marginalized}{fig:Results:constraints_galform_fiducial_xray}), this parameter space is strongly disfavoured.

The final important consideration folded into satellite count analyses such as ours is the estimated size of the total satellite galaxy population of the \MW{}. As we show in \appref{sec:Appendix:Structure_formation_constraints_N_sats} and \figref{fig:Results:constraints_galform_varying_NsatMW}, this can affect the stringency of the constraints significantly.
The true size of the \MW{} satellite galaxy population is unknown because most of the virial volume has not been surveyed. In recent years this figure has been updated several times in response to the influx of new data from several wide-area survey programmes and the application of different modelling techniques to account for incompleteness. Now, several estimates exist.
When we adopt the value of \NsatNum[MW]{183} favoured by \citet{nadler_milky_2020}, which disagrees with our earlier estimate of \NsatNum[MW]{124^{+40}_{-27}} \citep{newton_total_2018}, we rule out all parameterizations at and below \Ms[{\keV[7.1]}] using our fiducial \MW{} mass and parameterization of reionization. This is significantly less restrictive than the constraints obtained by \citet{nadler_constraints_2021}, who rule out all parameterizations of the \MSM{} below \Ms[{\keV[17.5]}] when assuming this number of satellite galaxies.
We are unable to account for the discrepancy between our results and theirs by changing the number of satellite galaxies alone.

Several components of the respective analyses could contribute to a discrepancy of this magnitude:
\begin{enumerate*}
    \item the width of the \Nsat[MW] uncertainty distribution,
    \item artificially disrupted subhaloes in simulations,
    \item the assumed mass of the \MW{} halo, and
    \item how the subhaloes are populated with galaxies.
\end{enumerate*}
We now discuss and compare each of these in turn.
\citet{newton_constraints_2021} demonstrated that the width of the distribution describing the uncertainty in the estimate of \Nsat[MW] affects the stringency of the constraints, and that broader distributions can weaken them by as much as \percent{15}. We use the \citet{newton_total_2018} probability distribution to model this uncertainty, which is slightly broader than the one used by \citet{nadler_constraints_2021}; however, the discrepancy is marginal so this is likely to be a subdominant effect.

\citet{newton_constraints_2021} also showed
that the artificial disruption of low-mass subhaloes caused by limited numerical resolution affects the constraints. Failing to account for this population, some of which could in principle host galaxies, 
artificially strengthens the constraints in analyses of this type.
\citet{nadler_constraints_2021} attempt to account for such artificially destroyed subhaloes,
finding that they comprise only \percent{10} of the whole subhalo population in gravity only zoom-in simulations of \MW{} mass haloes with a particle mass resolution of \Msunh[{3\times10^5}] \citep{mao_dependence_2015}.
This fraction is significantly less than the sevenfold increase in the number of subhaloes suggested by \citet[][in particular see \extfig{7}]{newton_constraints_2021} for similarly massive host haloes drawn from the \COCO{} simulations, which have a comparable mass resolution.
This discrepancy is a potentially significant factor in the disparity between our results and theirs.
It also highlights the considerable uncertainty that remains about the proper treatment of prematurely destroyed subhaloes.
For example, \citet{benson_tidal_2022} found that artificial disruption in the Caterpillar simulations reduces the host's subhalo population by \percent{10-20}. Most of the destruction is concentrated in the centre of the host, where as much as half of the subhaloes are destroyed by numerical instability. These results are consistent with the estimates used above.
A related issue concerns the modelling of the stellar component of prematurely destroyed subhaloes, which is poorly understood at present. It is unclear what fraction of the prematurely destroyed subhaloes would host an observable galaxy, and to what extent their stars would be affected by tidal stripping. It is thus an important systematic in satellite number count analyses.

We mentioned previously that the \MW{} halo mass adopted by \citet{nadler_constraints_2021} is \percent{20} larger than our fiducial assumption. As choosing a more massive \MW{} halo mass weakens their constraints slightly, we dismiss this as having only a negligible contribution to the discrepancy between the two sets of results.
Finally, as we showed in \secrefs{sec:Results:nuMSM_constraints:DM_only}{sec:Results:nuMSM_constraints:Galform}, the procedure used to populate \DM{} subhaloes with galaxies affects the constraints. \citet{nadler_constraints_2021} use subhalo abundance matching
to determine what fraction of the subhalo population hosts galaxies, whereas we adopt physically motivated semi-analytic prescriptions implemented in \Galform{}.

Subhalo abundance matching in the relevant mass regime has been shown to be inconsistent with the results of high resolution simulations by \citet{sawala_bent_2015}.
By contrast, \Galform{} provides a more nuanced physical portrayal of the complexities of galaxy formation on small scales, in particular, processes such as star formation and feedback; reionization; and gas heating and cooling; albeit with some simplifying assumptions.
Although \Galform{} is not calibrated to reproduce observational results at the scales of interest here, it nonetheless performs well when compared with observations of the abundance and radial profile of \MW{} satellite galaxies \citep{bose_imprint_2018,bose_little_2020} as well as with high-resolution hydrodynamical simulations that self-consistently track the evolution of low-mass \DM{} haloes and their baryonic components \citep[e.g.][]{grand_determining_2021}.
In comparison, the abundance matching approach suppresses the formation of faint galaxies more aggressively than we find when applying \Galform{}.
When combined with the possible under-correction for prematurely destroyed subhaloes, stronger constraints are an inevitable outcome.
We believe these two methodological differences are the root causes of the discrepancy between the results, although this can only be confirmed with a more detailed investigation and comparison. An analysis such as this is beyond the scope of this work and we leave it to other studies.

Our fiducial results alone do not exclude the \MSM{} as an explanation for the \keV[3.55] excess; however, they do significantly restrict the parameter space at \Ms[{\keV[7.1]}]. While X-ray non-detections already suggest that the \MSM{} cannot explain the \keV[3.55] excess, the lower limits are affected by uncertainties in modelling astrophysical backgrounds that could weaken these conclusions.
Further light will be shed on this question by \XRISM{} in the coming months. The \textit{Resolve} instrument onboard \XRISM{} has sufficient resolution to determine what flux originates from any emission line at \keV[3.55] compared to any background, and also enough resolution in principle to measure the width of such lines.
If these observations rule out all explanations within the realm of known physics, two possibilities remain: the \MSM{} sterile neutrino could account for only some fraction of the cosmic \DM{} abundance, or new physics are required to explain its origin.

\section{Conclusions}
\label{sec:Conclusions}
In this work we impose robust constraints on the parameter space of the \MSM{} by comparing predictions for the number of low-mass galaxies hosted by \MW{} mass \MSM{} haloes against recent estimates of the size of the \MW{} satellite galaxy population. Our results complement the updated constraints determined from X-ray non-detections, and using other astrophysical probes such as gravitational lensing, the \Lyman{\alpha} forest, and the modelling of gaps in stellar streams.

To determine our constraints we use the \EPS{} formalism
to produce reasonable estimates of the total abundance of substructure in \MW{} mass haloes (see \figref{fig:Methods:Calibrate_EPS}). We calibrate it to match gravity only simulations corrected to account for subhaloes artificially destroyed by numerical effects. This is important because omitting the prematurely destroyed subhaloes strengthens the constraints artificially \citep{newton_constraints_2021}.
We adopt two
distinct
approaches to populate the subhaloes with galaxies.
The first approach, in which we make no assumptions about galaxy formation physics and presume that all subhaloes host a galaxy (see \secref{sec:Results:nuMSM_constraints:DM_only}), produces the least stringent but most robust constraints.
This rules out, with at least \percent{95} confidence, all parameterizations of the \MSM{} with \MsConstraint{1.2} independently of the effective active--sterile neutrino mixing angle and the \MW{} halo mass (see \figref{fig:Results:constraints_eps_marginalized}).
When adopting the \citet{callingham_mass_2019} estimate of the \MW{} halo mass we exclude all parameterizations with \MsConstraint{1.4}.
These results are
consistent with
constraints derived from theoretical and observational analyses of the phase space density of \DM{} within \MW{} dwarf satellite galaxies \citep{boyarsky_lower_2009}.
In the latter case, the parameter space with \Ms[{\keV[7.1]}] is partially constrained at small mixing angles. If, as has been suggested recently, the \MW{} halo mass is at the lower end of estimates, the constraints at this rest mass strengthen somewhat. In combination with the lower limits determined from X-ray non-detections,
the parameter space at \Ms[{\keV[7.1]}] closes completely
in the albeit unlikely scenario in which the subhalo occupation fraction is extremely high. Adopting a more realistic estimate of the abundance of satellite galaxies in \MW{} mass haloes strengthens the constraints further.

The second approach, in which we apply the \Galform{} semi-analytic model of galaxy formation, incorporates physically motivated prescriptions of baryonic processes that suppress the formation of galaxies. In our analysis this renders more parameterizations of the \MSM{} incompatible with estimates of the \MW{} satellite count and tightens the constraint on the viable parameter space significantly.
One of the most important physical processes affecting the formation and evolution of low-mass galaxies is the reionization of hydrogen at early times.
Using \Galform{} we explore how changing the redshift at which reionization finishes, \zreion{}, and the circular velocity threshold, \Vcut{}, below which galaxy formation in \DM{} haloes is suppressed, affects the results (see \secref{sec:Results:nuMSM_constraints:Galform}). Adopting a fiducial parameterization of reionization with \zreion[7] and \Vcut[{\kms[30]}], we rule out with at least \percent{95} confidence all parameter combinations of the \MSM{} with \MsConstraint{4} independently of effective mixing angle and \MW{} halo mass (see \figref{fig:Results:constraints_galform_fiducial_xray}). The rest of the possible parameter space closes almost completely, and only the parameterizations of the \Ms[{\keV[7.1]}] \MSM{} with $0.4 \leq \sin^2\!\left(2 \mixAng{}\right)\, /\, 10^{-10} < 10$ remain viable.
All parameterizations of the \MSM{} with \Ms[{\keV[7.1]}] are excluded if $\MMW{}\leq\Msun[{8\times10^{11}}]$.
Consistent with previous work \citep{newton_constraints_2021}, we find that increasing \zreion{} and \Vcut{} strengthens the constraints, and the converse also holds (see \figref{fig:Results:constraints_galform_explore_reion}).

A key assumption in analyses such as ours is the size of the \MW{} satellite galaxy population, which is unknown because most of the virial volume has not been surveyed yet.
Our fiducial analysis adopts the \citet{newton_total_2018} estimate of the total population; however, several larger estimates have been published since then.
Assuming larger values of the \MW{} satellite count rules out larger fractions of the \MSM{} parameter space and strengthens our constraints somewhat (see \figref{fig:Results:constraints_galform_varying_NsatMW}). Notably, if \NsatNum[MW]{183} we rule out all parameterizations of the \MSM{} with \MsConstraint{7.1}, thus excluding the \MSM{} as the origin of the \keV[3.55] excess. However, compared to recent literature our constraints are less stringent.

We are unable to reproduce the strongest constraints reported by \citet{nadler_constraints_2021} even when we adopt a significantly larger number of \MW{} satellites.
We attribute the disparity between our results and theirs to both the different choice of particle physics calculation and to the differing methodologies used to populate the subhaloes with galaxies. The physically motivated semi-analytic prescription in \Galform{}, which we use, contrasts with the subhalo abundance matching adopted in \citet{nadler_constraints_2021}.
The latter approach, which has been shown to be inconsistent with the results of high resolution simulations by \citet{sawala_bent_2015}, suppresses the formation of faint galaxies more aggressively than the \Galform{} model.
A detailed and systematic comparison of the methodologies used in both studies, while undoubtedly of significant interest, is beyond the scope of this work and we leave this for future studies.
    
The recent launch of the JAXA \XRISM{} satellite \citep{terada_detailed_2021}, an almost like-for-like replacement of the ill-fated \Hitomi{} mission, will considerably improve observational capabilities at soft X-ray energies once commissioning is complete.
With higher sensitivity and resolution compared with previous instruments, \XRISM{} is poised to confirm or refute the suggestion that the \keV[3.55] excess corresponds to a detectable X-ray line.
This unconfirmed signal, which currently lacks a clear astrophysical origin, has motivated considerable theoretical and observational efforts to ascertain whether it may originate from the dark sector.
A successful confirmation of the signal and subsequent measurements of its width in various astrophysical systems could provide crucial evidence of the nature of the \DM{} \citep{lovell_anticipating_2023}. However, our results imply that it is unlikely to correspond to the sterile neutrino proposed in the \MSM{} framework.

\section*{Acknowledgements}
ON thanks Mariana Jaber for useful discussions.
ON acknowledges support from the
Polish National Science Centre under grant UMO-2020/39/B/ST9/03494.
ARJ and CSF were supported by STFC grant ST/L00075X/1.
CSF acknowledges support from the European Research Council~(ERC) through Advanced Investigator grant DMIDAS~(GA 786910).
MRL, SC, and JCH are supported by STFC grants ST/T000244/1 and ST/X001075/1.
This work used the DiRAC@Durham facility managed by the Institute for Computational Cosmology on behalf of the STFC DiRAC HPC Facility~(\url{www.dirac.ac.uk}). The equipment was funded by BEIS capital funding via STFC capital grants ST/K00042X/1, ST/P002293/1, ST/R002371/1 and ST/S002502/1, Durham University, and STFC operations grant ST/R000832/1. DiRAC is part of the National e-Infrastructure.

\textit{Software}: This research made use of \astropy{} \citep{the_astropy_collaboration_astropy_2013,the_astropy_collaboration_astropy_2018}, \matplotlib{} \citep{hunter_matplotlib_2007}, \numpy{} \citep{walt_numpy_2011,harris_array_2020}, \python{} \citep{van_rossum_python_2009}, and \scipy{} \citep{jones_scipy_2011,virtanen_scipy_2020}. This research also made use of the NASA Astrophysics Data System~(\url{http://adsabs.harvard.edu/}) and the arXiv e-print service (\url{http://arxiv.org/}). We thank their developers for maintaining them and making them freely available.

\section*{Data Availability}
The data used in this work are available upon reasonable request to the corresponding author.
A repository of reduced data and scripts to produce the figures in this manuscript will be made available on GitHub
\footnote{Supplementary materials: \github{Musical-Neutron/nuMSM_constraints}}
and will be archived in Zenodo.


\bibliographystyle{mnras}
\bibliography{archive} 



\appendix
\section{Dependence of structure formation constraints on the total number of Milky Way satellites}
\label{sec:Appendix:Structure_formation_constraints_N_sats}
\begin{figure}%
    \centering%
	\includegraphics[width=\columnwidth]{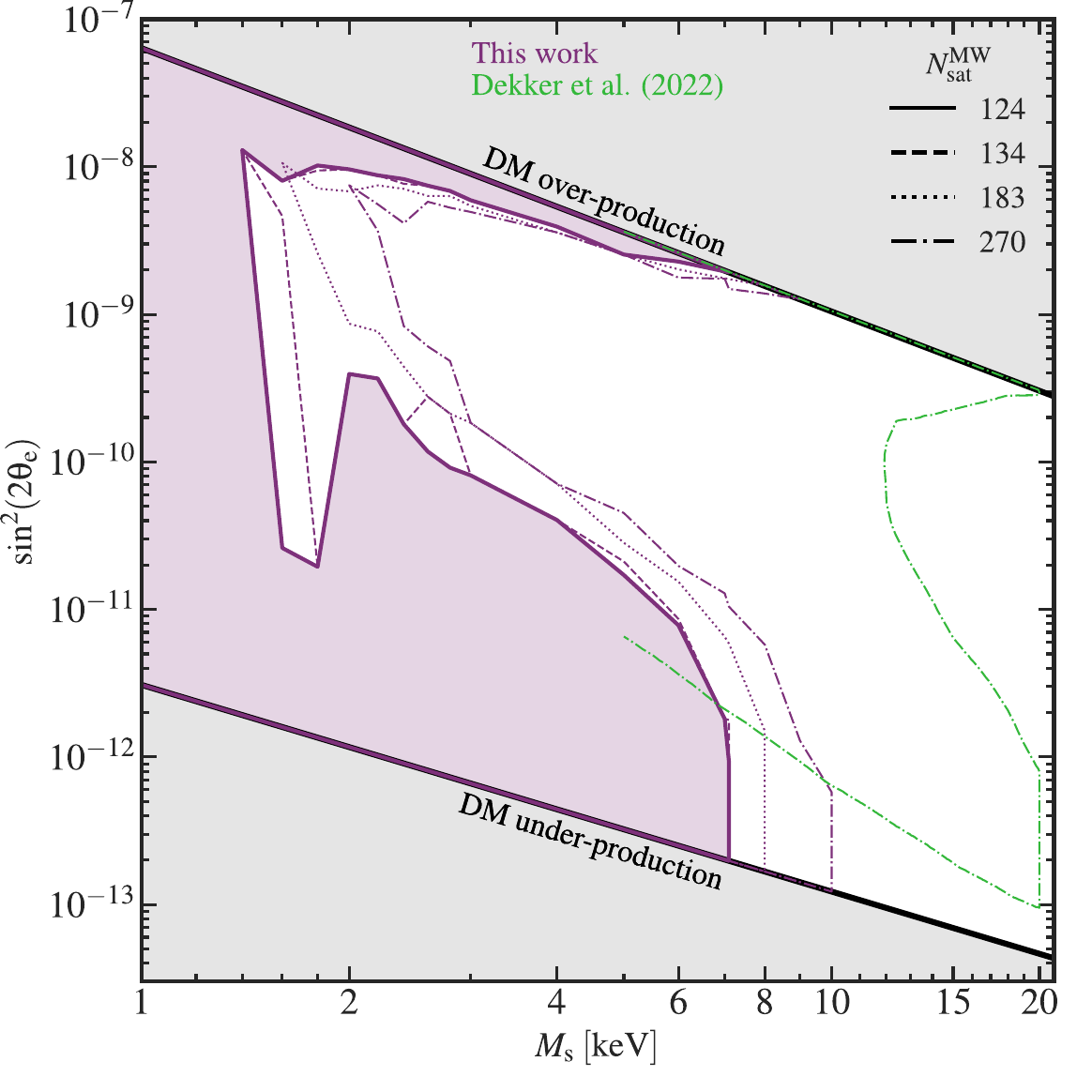}%
	\caption{Structure formation constraints on the \MSM{} parameter space from analyses of \MW{} satellite counts. As for several figures in the main text, the shaded regions in the upper and lower portions of the parameter space delimit parameter combinations for which the \DM{} is over- or under-produced if resonantly produced sterile neutrinos compose all of the \DM{} \citep{asaka_msm_2005,schneider_astrophysical_2016}. Our results are marginalized over the uncertainties in estimates of the \MW{} halo mass from \citet{callingham_mass_2019}. We compute constraint envelopes assuming a total \MW{} satellite galaxy population with a mean value of \NsatNum[MW]{124} (thick solid line), $134$ (dashed line), $183$ (dotted line), and $270$ (dot-dashed line). Parameter combinations within the envelopes are ruled out with at least \percent{95} confidence. For comparison we show the constraint envelope from \citet{dekker_warm_2022}.}%
	\label{fig:Appendix:Structure_formation_constraints_N_sat}%
	\vspace{-10pt}%
\end{figure}

The constraints obtained via satellite number counts are sensitive to the size of the Milky Way satellite galaxy population adopted for the analysis. This has not been measured yet because current surveys do not have sufficient depth to survey the entire \MW{} virial volume, nor do they cover the entire sky. In \figref{fig:Appendix:Structure_formation_constraints_N_sat}, we present the constraints we obtain by comparing \EPS{} predictions of the total abundance of substructure in \MW{} mass haloes with four different estimates of the size of the \MW{} satellite galaxy population, \NsatNum[MW]{} \citep{newton_total_2018,drlica-wagner_milky_2020,nadler_milky_2020}. As in \secref{sec:Results:nuMSM_constraints:DM_only}, we marginalize over the uncertainties in the \citet{callingham_mass_2019} estimate of the \MW{} halo mass. Our constraints become more restrictive as the assumed size of the \MW{} satellite galaxy population increases because this requires the parameterizations of the \MSM{} model to produce more substructure to remain viable. For completeness we also show our results when adopting \NsatNum[MW]{270} (dot-dashed line), which was reported by \citet{drlica-wagner_milky_2020} and was used in the analysis conducted by \citet{dekker_warm_2022}. Adopting this value of \Nsat[MW] produces very stringent constraints on the parameter space; however, \citet{nadler_milky_2020} noted that this estimate is inflated because the satellite distribution was assumed to be isotropic.

Alongside our constraints we plot the constraint envelope reported by \citet{dekker_warm_2022}, in which they use an \EPS{} framework and adopt the conservative assumption that all subhaloes host galaxies, as we did in \secref{sec:Results:nuMSM_constraints:DM_only}. As noted above, their adopted value of \NsatNum[MW]{270} is now recognised to be too large; however, the overall approach is similar to ours so we would expect our constraints obtained when assuming the same value to be comparable with theirs. As \figref{fig:Appendix:Structure_formation_constraints_N_sat} shows, our constraints are less restrictive, particularly where the effective mixing angle is large.
Four aspects of their analysis contribute to their stronger constraints.
First, their \EPS{} framework typically underestimates the number of subhaloes that survive until \z[0] by up to \percent{40} compared with the results of \Nbody{} simulations from \citet{lovell_properties_2014}. This discrepancy rises to \percent{60} in the \CDM{} model \citep[][\extapp{C}]{dekker_warm_2022}.
In part, this may be because they calibrate their \EPS{} framework by adopting values of the density field filter function parameters from \citet{schneider_structure_2015}.
The simulations used in that work do not account for subhaloes that were prematurely destroyed by numerical effects.
We showed in \citet{newton_constraints_2021} that underestimating the abundance of substructure by omitting the population of artificially destroyed subhaloes suppresses \fviable{} and can artificially strengthen constraints on the parameter space of \WDM{} models.

Secondly, their calculation of \fviable{} does not account fully for sources of uncertainty. The scatter in the subhalo mass function is modelled using a Poisson distribution, which is narrower than the negative binomial distribution favoured by \Nbody{} cosmological simulations \citep{boylan-kolchin_theres_2010}. Additionally, they do not account for the uncertainty in the estimate of the total \MW{} satellite galaxy population. Underestimating or omitting one or both of these sources of uncertainty can produce constraints that are too strict \citep[][\extsec{2.4}]{newton_constraints_2021}.

Third, they assume a \MW{} halo mass, \MMW{\Msun[10^{12}]}, which is \percent{20} less massive than the \citet{callingham_mass_2019} estimate that we use. Adopting smaller values of \MMW{} can increase the stringency of the constraints on the parameter space; however, as we showed in \secref{sec:Results:nuMSM_constraints:DM_only}, when assuming that all subhaloes are populated by galaxies the variation of the constraints as a function of \MMW{} is not significant. Therefore, we think that this is a subdominant contribution to the discrepancy between our results and those of \citeauthor{dekker_warm_2022} in \figref{fig:Appendix:Structure_formation_constraints_N_sat}.

Finally, in common with \citet{nadler_constraints_2021}, \citeauthor{dekker_warm_2022} adopt the \MSM{} momentum distributions computed by \citet{venumadhav_sterile_2016}. They are warmer than the momentum distributions we use, which complicates the comparison of our results with theirs.

\section{Modifications to the merger tree construction algorithm}
\label{sec:Appendix:Modifications_to_PCH}
In \citet[\extapp{B}]{newton_constraints_2021}, we noted that the Monte Carlo merger trees generated in \Galform{} using the standard implementation of the \citet{parkinson_generating_2008} algorithm overestimate the abundance of low-mass galaxies by a factor of two. We addressed this by applying an empirical correction to the theoretical galaxy luminosity function in which the satellite galaxy absolute magnitude, \MV{}, predicted by \Galform{} applied to the Monte Carlo merger trees is mapped to that predicted by \Galform{} applied to the \COCO{} simulation merger trees. This produced results in good agreement with the simulations; however, they are not self-consistent predictions of the \EPS{} methodology.

The discrepancy between the Monte Carlo and \Nbody{} predictions arises partly because the progenitor mass functions generated by the standard \citet{parkinson_generating_2008} algorithm under-predict those produced in cosmological simulations at the low-mass end.
This is because the \citet{parkinson_generating_2008} algorithm was calibrated against the \Millennium{} simulation, which cannot probe such low halo masses.
The \citet{parkinson_generating_2008} algorithm generates merger trees
probabilistically via a sequence of branching events in which the mass, \M{1}, in the halo at time, $t_1$, may be partitioned into two lower-mass progenitors with masses, $M_2$ and $\left(M_1 - M_2\right)$, at an earlier time, $t_2$. A complete merger tree is built up by repeating this process for each progenitor halo at each successive timestep. The size of the timestep, \dv{t}, is selected to be sufficiently small such that the halo is unlikely to have more than two progenitors above some mass resolution threshold at the earlier time.
The rate of progenitor formation is controlled by the binary merger branching rate,
\begin{equation}
    \label{eq:Appendix:PCH_tree_branching:binary_branching_rate}
    \begin{split}
        \fdv{p}{\omega} = \int^{M/2}_{M_{\rm min}}
        \frac{M_1}{M'}
        &\fdv{f\!\left(M'\right)}{t}
        \fdv{S\!\left(M'\right)}{M'}
        \left|\fdv{t}{\omega}\right|\\
        &\quad \times G\left[\omega,\, \sigma\!\left(M_1\right),\, \sigma\!\left(M'\right)\right]\, \dv{M'}\,,
    \end{split}
\end{equation}
where $\omega = \delta_{\rm c,\, 0}\, /\, D\!\left(t\right)$ is the ratio of the linear theory critical overdensity threshold for gravitational collapse in \CDM{} models, $\delta_{\rm c,\, 0},$ to the linear growth-rate factor, $D\!\left(t\right)$; $M_{\rm min}$ is the required minimum mass resolution of the merger tree; $\dv{f}\, /\, \dv{t}$ is the first-crossing rate distribution, the rate at which the excursion set trajectories first cross the overdensity threshold \citep{bond_excursion_1991}; $\SigmaFunc{M}=\sqrt{S\!\left(M\right)}$ is the square root of the variance, \textit{S}, of the matter density field;
and $G\left[\omega,\, \sigma\!\left(M\right),\, \sigma\!\left(M'\right)\right]$ is an empirical modification introduced to obtain results consistent with those from cosmological \CDM{} simulations.
In the standard implementation of the \citet{parkinson_generating_2008} algorithm \textit{G} is given by
\begin{equation}
    \label{eq:Appendix:PCH_tree_branching:standard_G}
    \begin{split}
        G\left[\omega,\, \SigmaFunc{M_1},\, \SigmaFunc{M_2}\right] =
        G_0
        \left[\frac{\SigmaFunc{M_2}}{\SigmaFunc{M_1}}\right]^{\gamma_1}
        \left[\frac{\omega}{\SigmaFunc{M_1}}\right]^{\gamma_2}\,,
    \end{split}
\end{equation}
which introduces the free parameters, $G_0,\, \gamma_1,$ and $\gamma_2$ that are calibrated using \DM{}-only cosmological simulations \citep[see e.g.][\exttab{1}]{benson_halo_2019}.

As we discussed above, this implementation of the algorithm over-predicts the number of low-mass subhaloes because the empirical modification underestimates the simulated progenitor mass functions.
We attempt to address this by introducing an explicit dependence on $D\!\left(t\right)$ and the logarithmic slope of the \SigmaFunc{M} relation such that
\begin{equation}
    \label{eq:Appendix:PCH_tree_branching:modified_G}
    \begin{split}
        G\left[\omega,\, \right.&\left.\SigmaFunc{M_1},\, \SigmaFunc{M_2}\right] = G_0
        \left[\frac{\SigmaFunc{M_2}}{\SigmaFunc{M_1}}\right]^{\gamma_1}
        \left[\frac{\omega}{\SigmaFunc{M_1}}\right]^{\gamma_2}\\
        &\quad \times \left\{1-\left[\frac{\SigmaFunc{M_2}}{\SigmaFunc{M_1}}\right]^2\right\}^{\gamma_3}
        {D\!\left(t_2\right)}^{\gamma_4}
        \left|\fdv{\ln \SigmaFunc{M_2}}{\ln M}\right|^{\gamma_5}\,.
    \end{split}
\end{equation}
This paramterization was chosen to ensure that, if applied to a scale-free cosmology, the merger trees would satisfy the required self-similarity constraint \citep{efstathiou_gravitational_1988}.
This introduces a further three new free parameters: $\gamma_3,\, \gamma_4,$ and $\gamma_5$ that are also determined via calibration to \Nbody{} simulations. Throughout this work, we use the latter form of \textit{G} with $G_0=0.5024,\, \gamma_1=0.2493,\, \gamma_2=-0.095846,\, \gamma_3=0.045166,\, \gamma_4=0.2342,$ and $\gamma_5=-0.1878$, based on a preliminary recalibration to recent simulations (A.~J.~Benson et al., in preparation).


\bsp	
\label{lastpage}
\end{document}